\documentclass[manuscript,nonacm]{acmart}
\usepackage{multirow}        % loaded once (was duplicated)
\usepackage{dirtytalk}
\usepackage{comment}
\usepackage{enumitem}
\usepackage{tabularx,makecell,booktabs,array}
\usepackage{longtable}       % moved here from conflicting block below
\usepackage{ragged2e}        % moved here; needed for P column type
\usepackage{pdflscape}       % for landscape pages in appendix longtable
\usepackage{placeins}   % in the preamble

\definecolor{mygooglecolor}{RGB}{216, 191, 216}
\definecolor{mychatcolor}{RGB}{255, 192, 203}
\definecolor{mygencolor}{RGB}{173,216,230}
\definecolor{dodgerblue}{rgb}{0.12, 0.56, 1.0}
\definecolor{green(ncs)}{rgb}{0.0, 0.62, 0.42}
\definecolor{myblue}{HTML}{00008B}

\newcolumntype{C}{>{\centering\arraybackslash}X}  % centered X column (tabularx)
\newcolumntype{P}[1]{>{\RaggedRight\arraybackslash}p{#1}} % moved here from body
\AtBeginDocument{%
  }

\setcopyright{acmlicensed}
\copyrightyear{2018}
\acmYear{2018}
\acmDOI{XXXXXXX.XXXXXXX}
\acmConference[Conference acronym 'XX]{Make sure to enter the correct
  conference title from your rights confirmation email}{June 03--05,
  2018}{Woodstock, NY}
\acmISBN{978-1-4503-XXXX-X/2018/06}

\definecolor{myblue}{HTML}{00008B}
\begin{document}

\title[AI as Anchor]{AI as Anchor: How Interview Sequencing Relates to the Appropriation of Synthetic Stakeholder Voices in Design Process}

%\author{Blinded}

%\begin{comment}
%\titlenote{Accepted to the International Journal of Human-Computer Interaction, November 2026}
\author{Osnat Mokryn}
\authornote{Corresponding author}
\email{omokryn@is.haifa.ac.il}
\orcid{0000-0002-1241-9015}
\author{Oshrat Ayalon}
%\email{oayalon@is.haifa.ac.il}
\author{Noga Dines}
%\email{}
\affiliation{%
  \institution{Information Systems, University of Haifa}
  \city{Haifa}
  \country{Israel}
}

\author{Orit Shaer}
\affiliation{%
  \institution{Computer Science, Wellesley College}
  \city{Wellesley}
  \state{MA}
  \country{USA}}
%\email{}

%\renewcommand{\shortauthors}{Mokryn et al.}
%\end{comment}

\begin{abstract}
Designers increasingly use large language models to simulate stakeholder 
conversations. Yet it remains unclear how AI Simulated Stakeholders (AIS) are taken up when used alongside human stakeholder engagement. We report a field study in a single-institution, single-semester design course in which 8 teams (N=21) constructed, evaluated, and peer-reviewed an initial AIS. Then, the teams interviewed both AIS and human stakeholders in different sequences: AIS-interview-first or human-interview-first. Drawing on interview transcripts, design artifacts, course-embedded reflections, and post-course qualitative sessions, we examine exploratory patterns associated with these sequencing conditions.

We find that students did not describe AIS as a neutral addition to the process. Instead, the order in which AIS and human stakeholders were formally interviewed was associated with different ways of interpreting and using information from the two sources. When the AIS interview came first, it provided a broad but often general frame that students later reconsidered through human interviews. When the AIS interview followed human engagement, AIS was more often used to revisit and validate insights already grounded in lived experience. Across both interview orders, students described AIS as a point of reference, or “anchor,” but its reported role varied with its position in the formal interview sequence. In the final designs, a larger proportion of prominent human-stakeholder-derived concerns than AIS-derived concerns had corresponding design elements.

We contribute: (1) a characterization of AIS as a reference point whose reported role varies with its position in the formal interview sequence, (2) exploratory empirical evidence that the implemented interview-order conditions were associated with distinct patterns of interpreting and using stakeholder input, and (3) implications for integrating AIS into design workflows in ways that preserve direct engagement with human stakeholders.
%We find that AIS did not function as a neutral addition to the process. Instead, the 
%sequence in which AIS was introduced was associated with different ways of interpreting and using subsequent information. When used first, AIS provided a broad but often general frame that students later refined through human interviews. When used after human engagement, AIS was primarily used to revisit and validate insights that were already grounded in lived experience. Across both sequences, students described AIS as an initial point of reference, ``anchor'', but its role shifted with timing. In final designs, insights derived from human stakeholders were incorporated more prominently than those derived from AIS.

%We contribute: (1) a characterization of AIS as a reference point whose role depends on its position within the design process, (2) exploratory evidence that sequencing is associated with distinct patterns of interpreting and using stakeholder input, and (3) implications for integrating AIS into design workflows while preserving direct  human stakeholders engagement.
\end{abstract}



\maketitle

\section{Introduction}
Sociotechnical systems represent complex configurations where technology, 
human actors, and social structures interact dynamically, creating emergent 
behaviors that cannot be predicted from analyzing individual components 
in isolation \cite{berg1999patient, friedman2013value}. Designing such systems 
requires engaging with the perspectives of those affected by them, raising 
questions about whose interests are represented and whose voices are heard. 
Participatory design addresses this by involving stakeholders directly in 
design processes, while value-sensitive design provides methods for eliciting 
and embedding stakeholder values into technical decisions \cite{schuler1993participatory, friedman2013value}. 

In practice, however, participatory approaches face constraints. Limited time, 
resources, and access often restrict designers to engaging only a small number 
of stakeholders, providing a partial and sometimes unbalanced view of the design space.

Recent advances in conversational AI introduce new possibilities. Large language 
models can generate interactive agents that simulate stakeholder dialogue. Prior 
work has examined such systems either as substitutes for human participants or as 
standalone design aids, with mixed findings regarding their realism, usefulness, 
and limitations \cite{agnew2024illusion, kapania2025simulacrum, anthis2025llm, smrke2025exploring}. 
Less is known about how such agents function when used \textit{alongside} authentic stakeholder 
engagement within an ongoing design process.

A key question is therefore, not only whether AI-generated stakeholders are useful, but how 
their introduction into the process relates to the interpretation of subsequent input from authentic stakeholders. In this study, all teams constructed and evaluated an AIS before sequence assignment. Sequencing, therefore, refers specifically to the order of the two formal interviews, rather than to initial exposure versus non-exposure to AIS. 
When designers engage with different sources of information over time, interview order may be associated with what is noticed, trusted, 
and carried forward into design decisions. This relationship remains underexplored in hybrid human-AI stakeholder engagement.

In this paper, we examine \textbf{AI Simulated Stakeholders (AIS)}, LLM-generated 
agents designed to emulate concrete stakeholder roles through conversational interaction. 
Rather than treating AIS as replacements for human stakeholder participation, we study how they are 
used alongside human stakeholders and how their position within the design process relates 
to designers' interpretation and use of stakeholder input.

We investigate this question in a naturalistic field study conducted in a mixed-level 
design course. Eight teams (N=21) constructed AIS before sequence assignment and then interviewed AIS and human stakeholders in one of two formal interview orders: AIS-interview-first or human-interview-first, as part of a 
semester-long project. This setting allows us to examine how students who are novice designers navigate 
synthetic and authentic voices over time and how concerns from each source are incorporated 
into evolving design work.

Here, we use appropriation to refer to how students take up AIS in practice: constructing and adapting it, interpreting and evaluating its outputs, validating or challenging them, and integrating, repurposing, or rejecting them within the design process.

We ask:

\noindent\textbf{RQ1}: How do novice designers appropriate AIS during early-stage design work? 

\noindent\textbf{RQ2}: How does the order of the formal AIS and human stakeholder interviews relate to students’ expectations, interpretations, and interview practices?

\noindent\textbf{RQ3}: How are insights from AIS and human stakeholders reflected in the design decisions students produce?

%We treat the interview sequence as an analytic lens for examining how the timing of AIS use relates to how students interpret stakeholder input and integrate it into design decisions. 

While the study is situated in a university design studio, we approach this 
setting as a field site for observing how novice designers engage with synthetic and authentic 
voices, rather than as an evaluation of educational outcomes.

\noindent\textbf{This study makes the following contributions:}
\begin{enumerate}[label=\arabic*.]
    \item \textbf{AIS as a reference point:} We characterize how students used AI Simulated Stakeholders as provisional or secondary reference points across two  interview orders.

    \item \textbf{Interview order and interpretation:} We provide exploratory evidence that the implemented sequencing conditions were associated with different reported ways of interpreting and using stakeholder input.

    \item \textbf{Implications for design practice:} We outline considerations for integrating AIS into design workflows in ways that support engagement with human stakeholders.
\end{enumerate}

The remainder of this paper is organized as follows. Section 2 reviews related work. Section 3 describes the study setting, participants, and data collection. Section 4 outlines the AIS construction process. Section 5 presents our findings. Section 6 discusses implications, and Section 7 concludes.

\section{Related Work}
We examine AI Simulated Stakeholders (AIS) as one instance of AI-mediated, synthetic stakeholder representations that are woven into early-stage design work, alongside traditional stakeholder engagement in sociotechnical systems. In here, we focus on how novice designers appropriate AIS in practice, how they navigate synthetic and authentic voices over time, and how these engagements relate to the design decisions they ultimately produce.

This study examines how novice designers learn to engage stakeholder voices in sociotechnical design contexts where participatory methods are central to practice. We draw on research in design empathy and personas, emerging work on synthetic users and AI-generated personas, and studies of human–AI workflows in design education.

\subsection{Stakeholder Engagement in Sociotechnical Design Education}

Design students learning to work on sociotechnical systems are typically taught to engage diverse stakeholders directly, drawing on participatory design and value-sensitive design traditions.
Participatory design emerged from Scandinavian workplace democracy movements, 
establishing the principle that those affected by design should participate 
in the design process \cite{bjerknes1987computers, ehn1988work}. This tradition 
emphasizes direct involvement of the various stakeholders that represent various user groups needs, while making sure to also 
include underrepresented voices~\cite{schuler1993participatory}. The disability 
rights movement's principle of ``Nothing About Us Without Us'' \cite{charlton1998nothing} 
further reinforces that authentic participation cannot be substituted by designer 
imagination or proxies.

Authentic stakeholder voices are central to participatory design because they surface lived experience, contextual nuance, and power asymmetries that are often flattened or abstracted in designer-generated representations \cite{friedman2019value,schuler1993participatory,kouprie2009framework}.

Value-centered design examines how technologies shape and are shaped by values 
and moral concerns~\cite{friedman2013value}. As moral concerns and values are 
embedded within the social aspects of organizations, they are often invisible 
to designers. Value design relies on direct engagement with affected 
communities to ground requirements, negotiate trade-offs, and justify design 
choices in lived experience rather than abstractions~\cite{friedman2019value}.

In practice, however, real-world projects operate under time and budget limitations 
that restrict the number of stakeholders who can be meaningfully engaged. Delgado 
et al.~\cite{delgado2023participatory} found a tension between aspirations for empowerment and practical constraints 
such as resources, timelines, and expertise, leading to the adoption of algorithmic 
proxies that serve as user stand-ins. This is particularly problematic when 
stakeholder groups include marginalized or hard-to-reach populations who may 
be geographically dispersed, have limited availability (e.g., doctors, lawyers, 
etc.), or face barriers to participation~\cite{moser2012revisiting,guan2024leveraging}.

\subsection{Design Empathy and Personas}
Personas are fictional characters that represent human stakeholder segments~\cite{cooper1999inmates}. Personas help designers form a mental model of users, enabling them to anticipate 
behavior in scenarios while avoiding projecting their own needs onto the design~\cite{nielsen2013personas}.  Personas' success lies in that people remember individuals better than abstractions, 
empathize with characters in stories, and adopt perspectives more readily when 
given a concrete, humanized anchor~\cite{grudin2006personas}.

A key benefit is that personas enhance designer empathy and understanding 
of stakeholders~\cite{pruitt2003personas,miaskiewicz2011personas}.
Grudin~\cite{grudin2006personas} describes design empathy as operating through 
sophisticated cognitive mechanisms rooted in the fundamental human ability to 
attribute mental states to others, theory of mind. People ``find it natural 
to create and use models of other people'', drawing on basic cognitive abilities 
to interact in complex social environments~\cite{grudin2006personas}. In design 
contexts, this translates into designers using personas to construct detailed 
mental models of user experiences, motivations, and constraints.  Empathy involves 
affective matching, self-oriented perspective-taking, and other-oriented perspective-taking~\cite{amy2011understanding}. 
Kouprie and Visser~\cite{kouprie2009framework} distinguish levels of design 
empathy ranging from superficial understanding to embodied experience that 
might be achieved when the persona's design is rich and detailed.

Prior work has shown that personas, while widely used, often abstract away contextual nuance and can foreground designer assumptions rather than stakeholders’ lived experiences~\cite{marsden2016stereotypes,salminen2023choice,csengun2024there}.

 Personas are meant to represent users who cannot participate directly, 
yet the act of representation without participation risks epistemic injustice 
\cite{fricker2007epistemic}. Personas, as representations of users, inevitably 
make value judgments about which characteristics matter and whose voices are 
centered. Hence, Costanza-Chock~\cite{costanzachock2020design} 
argues that design justice requires moving beyond representation to actual 
participation by affected communities.

\subsection{Algorithmic Proxies and AI-Generated Stakeholder Representations}

As participatory design faces practical constraints~\cite{guan2024leveraging}, researchers and practitioners 
have increasingly turned to computational representations that stand in for 
human participants. These simulated proxies take several forms, each with 
distinct characteristics and applications.

\textbf{Algorithmic proxies} broadly encompass computational representations 
that encode human needs or behaviors through data-driven models rather than 
direct engagement~\cite{holzinger2022personas}. Delgado et al.~\cite{delgado2023participatory} demonstrate that practitioners increasingly adopt proxy-based tactics as pragmatic compromises between participatory ideals and real-world constraints of time, resources, and stakeholder availability.  Information-driven personas, termed ``imaginary people representing real numbers'' were created through computational analysis of online behavior patterns and social media mining~\cite{an2018imaginary,jung2018automatic}.  Salminen et al.~\cite{salminen2022use}, in their review of the use of personas as algorithm proxies, show that data-driven personas have proliferated across domains from healthcare to urban planning as scalable alternatives to traditional ethnographic methods. Recent research has begun to examine the usability of AI-generated personas themselves. Kapania et al.~\cite{kapania2025simulacrum} 
demonstrate that when large language models are used as research participant proxies, 
they create what they term ``surrogate effects,'' which raises epistemological concerns 
about whether these models can represent lived experience, social agency, or 
the dynamic and contested nature of stakeholder needs. Kaate et al.~\cite{kaate2025fourth} compared chat-based and profile personas, identifying issues such as verbosity, information overload, and the so-called ``fourth wall'' effect, where users felt they were conversing with a description of a persona rather than the persona itself. Their work highlights how different presentation formats influence user perceptions and task performance. In contrast, our study investigates how and when to integrate the AI Simulated Stakeholders (AIS) in design processes.

\textbf{AI-generated personas} represent a specific application where large 
language models are prompted to produce enhanced character profiles for design 
exploration. Goel et al.~\cite{goel2023preparing} examined how novice designers 
could use GPT-3 to assist in persona creation, finding that while AI-generated 
personas were comparable to manually created ones, they were rated lower on 
credibility, empathy, and memorability. Their work focused primarily on using 
AI to elaborate and refine static persona documents rather than enabling conversational 
interaction.

\textbf{LLM social simulations} deploy multiple synthetic agents to replicate 
populations or communities for large-scale research purposes. Anthis et al.~\cite{anthis2025llm} 
 examine the fundamental challenges of using LLMs as human research 
proxies and identify five recurring issues: \emph{diversity} (limited representation 
of demographic variety), \emph{bias} (systematic distortions toward majority 
perspectives), \emph{sycophancy} (tendency to agree with users), \emph{alienness} 
(responses that feel artificial or non-human), and \emph{generalization} (poor 
performance across different contexts).  Smrke et al.~\cite{smrke2025exploring} compared 
LLM-generated personas with human-crafted personas in obesity research contexts, 
finding that LLM-generated personas were generally perceived as believable, 
relatable, and informative, with some even outperforming human-crafted personas 
on empathy and likability dimensions.

\textbf{AI Simulated Stakeholders (AIS)} represent a distinct approach within 
this landscape. Unlike static AI-generated personas, AIS engage in real-time dialogue while maintaining specific stakeholder 
identities. Unlike population-level social simulations, AIS instantiate individual 
stakeholder roles within specific sociotechnical contexts, enabling designers 
to conduct one-on-one conversational interviews. Most critically, AIS are designed 
not as replacements for human participation, but as augmentation tools that 
can supplement limited stakeholder access while preparing designers for authentic 
engagement. 

The introduction of generative AI in design practice raises questions about 
whether AI should augment human capabilities or substitute for human participation 
entirely~\cite{raisch2021artificial}. This distinction is particularly relevant 
when stakeholder access is constrained.

Wright and McCarthy~\cite{wright2008empathy} emphasize the role of empathy in 
designer–user relations; it remains unclear whether AI-mediated interactions 
can cultivate similar levels of empathic understanding as traditional human-centered 
methods. Traditional personas build empathy through the research and synthesis 
process itself~\cite{grudin2006personas}, where designers must actively construct 
mental models from user data. AIS bypass this constructive process, potentially 
limiting empathy development despite offering conversational interaction. Recent work on AI-simulated user agents has examined whether structured prompting can improve the quality of synthetic user representations~\cite{fan2026simulated}. Yet improving the quality of a simulated representation does not resolve the question of how designers interpret and prioritize its contributions relative to those of human stakeholders. Mapping AI-based design support systems onto the Double Diamond model, Lee et al.~\cite{lee2025when} found that existing systems predominantly support solution generation rather than problem exploration and definition. This distinction is particularly consequential in stakeholder research, where AI-generated information may shape designers' understanding of the problem before potential solutions are considered.

In design education contexts, students face similar constraints to practitioners, such as limited time, access to stakeholders, and inexperience in conducting stakeholder research. This makes design courses a particularly relevant site for examining how novice designers learn to navigate stakeholder engagement practices and appropriate new AI tools within those practices.

\subsection{Human-AI Workflows and Timing in Design Practice}

Recent work on human-AI collaboration in design has begun to treat AI tools not as isolated interventions, but as part of evolving workflows that unfold over time. Shin et al.~\cite{shin2024understanding} studied human-AI workflows for generating personas, and showed that how tasks are divided between humans and LLMs (e.g., who identifies key characteristics, who groups data, who generates the final persona) shapes both the quality of the output and the designer’s experience. This work frames persona generation as a collaborative process rather than a one-off AI output, which resonates with our interest in how students appropriate AIS over the course of a semester-long project.

In design education, there is growing attention to how students learn to work with generative AI as a collaborator or ``virtual colleague'' in early design stages. Goel et al.~\cite{goel2023preparing} investigate how novice designers use AI text-generation tools to support persona creation in a course setting, finding that while AI-generated personas were comparable to manually created ones, they were rated lower on \emph{credibility, empathy, and memorability}. Interestingly, Jung et al.~\cite{jung2025personacraft} integrated LLMs with survey data analysis to generate personas that ``humanize'' data segments. Their PersonaCraft system scored highly on \emph{clarity, completeness, fluency, consistency, and credibility} in multi-stakeholder evaluations. 

In design studies, Sattele and Ortiz~\cite{sattele2024generating} took a critical qualitative approach by conducting a Mexico City case study that revealed cultural biases and stereotyping risks in AI-generated personas, while proposing scalability benefits for idea generation.  Gu et al.~\cite{gu2025synthetic} found that designers who were using \emph{persona-based Synthetic Users} during single-session ideation engaged in \emph{feedback-seeking, assumption-testing, and scenario role-playing} types of interactions, that are impossible with static personas. Our naturalistic study extends this to AIS within complete semester-long design processes, examining how timing shapes interaction patterns across expectations, interview practices, and design traceability.

However, most of this work focuses on a single use of AI personas or synthetic users within a bounded activity (e.g., a workshop or a single assignment). There is less empirical work that follows novice designers over time in a naturalistic, course-embedded project, examining how the \emph{timing} of AI use, when in the design process AIS are introduced, shapes students’ expectations, interview practices, and how insights from synthetic and authentic voices are reflected in their design decisions. Our study contributes to this emerging conversation by treating the sequence of AIS and human stakeholder interviews not as an experimental manipulation, but as a lens into how students navigate and differentiate between synthetic and authentic voices in early-stage design work.
\subsection{Research Gap}

While recent literature has identified fundamental limitations in LLM social simulations~\cite{anthis2025llm,smrke2025exploring,kaate2025fourth}, little is known about how design students appropriate AIS as part of their learning process, especially when conducting stakeholder research for the first time in semester-long projects. Most prior work examines controlled tasks or single-session workshops, whereas our study observes how students navigate synthetic and authentic voices over time in a naturalistic classroom setting.

In particular, few empirical studies have examined how the \emph{timing} of AIS use shapes students’ expectations, interpretations, and interview practices in a real design course. It remains unclear how introducing AIS early versus later in the process affects how students prepare for and conduct interviews with human stakeholders, and how they interpret and integrate insights from each.

Similarly, there is limited understanding of how insights derived from AIS and from human stakeholders are reflected in the design decisions students produce. Our study addresses this gap by presenting a naturalistic case study of how novice designers appropriate AIS during early-stage design work, how the timing of AIS use shapes their practices, and how insights from synthetic and authentic voices are traced into their design artifacts.

\section{Study Context and Methods}
\subsection{Course Setting and Pedagogical Structure}
The study was conducted as part of a semester-long university course on sociotechnical system design. Participants were undergraduate and graduate students with varied technical and design experience, representing multiple gender identities and diverse ethnic backgrounds.

\paragraph{Course structure and timeline.}
The study was embedded in a 12-week flipped classroom course on ethical design of sociotechnical systems with generative AI. Each week typically combined three components: pre-class preparation, including an assigned reading and a short written response; an in-class discussion connecting the reading to sociotechnical design; and a hands-on exercise or project activity. The semester-long group project developed incrementally alongside these activities, with students submitting intermediate milestones and receiving feedback throughout the course. Before any group could recruit or interact with human participants, all members were also required to complete human-subject research training.

Weeks 1–3 introduced sociotechnical systems and stakeholder perspectives. Students first experimented with LLMs in short in-class exercises and discussed Berg’s~\cite{berg1999patient} sociotechnical analysis of healthcare information systems, using it to examine how technical systems interact with organizational practices and human actors. The following week focused explicitly on identifying stakeholders and understanding how different actors experience the same system, supported by Beede et al.’s~\cite{beede2020human} human-centered evaluation of an AI system deployed in clinical practice. During this period, students introduced themselves, formed groups of two to three, and began identifying possible project domains.

Weeks 4–6 shifted toward generative AI, design cognition, and AI-supported persona creation. Students discussed the metacognitive demands of generative AI using Tankelevitch et al.~\cite{tankelevitch2024metacognitive} and practiced constructing more complex prompts using the Six Thinking Hats technique~\cite{debono1985six}. They then studied stakeholder persona construction, informed by Goel et al.’s~\cite{goel2023preparing} work on human-AI collaboration in persona creation, and participated in a guest lecture and hands-on persona-building exercise. By the end of this phase, groups had finalized their project ideas.

Weeks 7–8 moved from project definition to stakeholder-engagement planning. Students examined an example of an HCI intervention and empirical evaluation~\cite{huh2023genassist}, and then used Friedman and Hendry’s Envisioning Cards~\cite{friedman2012envisioning} to consider stakeholders, values, and possible downstream consequences of their proposed systems. During this period, groups planned their human and AIS interviews, finalized their stakeholder roles, and were assigned to the two interview sequences.

Weeks 8–12 combined the interview and design work with broader ethical reflection. Class topics included dark patterns, values work and influence within organizations, and unintended consequences of technology. For example, students discussed Bongard-Blanchy et al.’s work on users’ experiences of dark patterns~\cite{bongard2021definitely}, Wong’s analysis~\cite{wong2021tactics} of how UX professionals engage in values work without formal authority, and Veinot et al.’s account of how well-intentioned informatics interventions can exacerbate inequality~\cite{veinot2018good}. Alongside these discussions, groups worked on successive project milestones, received in-class feedback, synthesized their stakeholder interviews, developed their design interventions, and presented their final projects in the final week

During the joint preparation phase, all eight groups constructed an initial AIS, evaluated and peer-reviewed it, and developed interview protocols for both the AIS and human stakeholder. Every team had therefore already encountered and worked with AIS before the sequence assignment. After this shared preparation, groups were randomly assigned to one of two formal interview orders: four teams interviewed their AIS and then a human stakeholder, while four interviewed a human stakeholder and then their AIS. We refer to these conditions as AIS-interview-first and human-interview-first, respectively. Groups were permitted, but not required, to adapt their protocols or AIS between interviews. %Figure 1 illustrates this structure.
\begin{figure}
    \centering
\includegraphics[width=0.95\linewidth]{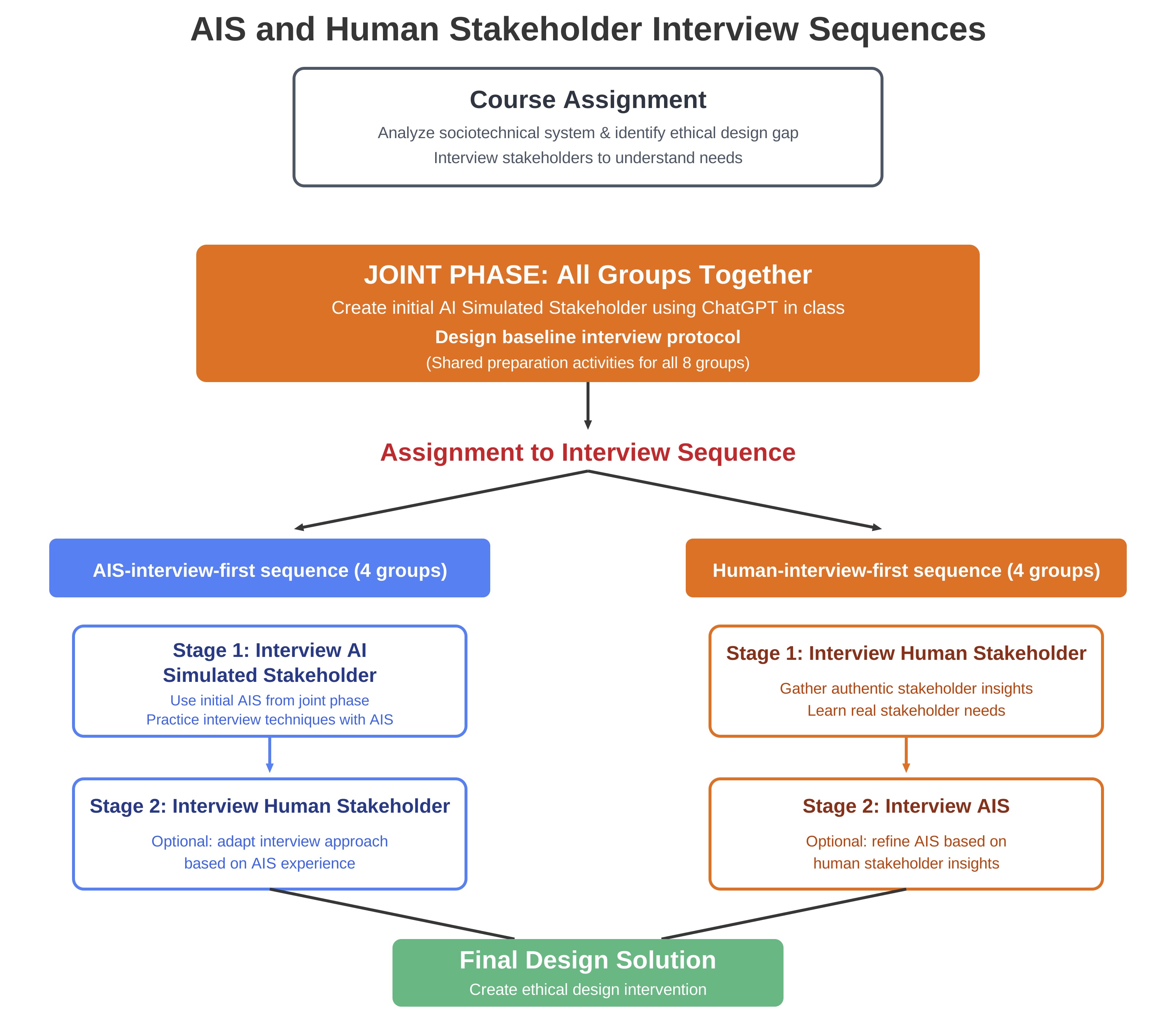}
   \caption{\small{\textbf{Study Structure.}  All teams first constructed, evaluated, and peer-reviewed an initial AI Simulated Stakeholder and developed interview protocols during shared preparation. They were then assigned to one of two formal interview orders: AIS-interview-first or human-interview-first. Groups were permitted, but not required, to adapt the AIS or interview protocol between interviews. The conditions therefore represent formal interview order together with any discretionary adaptation; they do not represent initial AIS exposure versus non-exposure.}}
    \label{fig:exp}
\end{figure}
Figure~\ref{fig:exp} illustrates the study structure. The study was embedded in a flipped classroom on ethical design for sociotechnical systems with generative AI, where students had previously engaged with key readings and experimented with AI tools in short in-class exercises. During this initial phase, student groups were formed and each chose a sociotechnical system along with the ethical aspect they wished to investigate, as detailed in Table~\ref{tbl:projects}.

The process then moved into a joint preparation phase (highlighted in orange) in class, during which each of the eight groups constructed their initial AIS, an LLM-generated conversational agent intentionally designed to emulate a concrete stakeholder role. Each group prompted the LLM to create an AIS that would have a similar role to their chosen stakeholder.  Most students used the free version of ChatGPT (GPT-4o) available at the time via the web interface~\cite{openai2024chatgpt}. Yet, they were free to use various LLMs and some tried out several. 
The AIS creation process is detailed in Section~\ref{sec:ais}. Following this joint preparation phase, all groups had developed an initial AIS and were ready to conduct stakeholder interviews. Groups then drafted interview protocols for both their human stakeholder and their AIS, which were reviewed and finalized with the course instructor before interviews began. 

Next, to examine how formal interview order related to students’ practices and perceptions, groups were assigned to one of the two interview-order conditions shown in Figure~\ref{fig:exp}. Four groups formally interviewed their AIS before interviewing a human stakeholder, and four interviewed a human stakeholder before formally interviewing their AIS. Assignment was conducted using a random-number generator and was independent of project domain or team composition. Because teams selected their project topics before assignment, project domain and stakeholder characteristics were not matched or balanced between conditions. Moreover, human-interview-first teams could use information from the human interview when optionally refining their AIS, whereas AIS-interview-first teams could use their AIS interview when optionally adapting their human-stakeholder protocol. The conditions therefore capture formal interview order together with any discretionary adaptation undertaken between interviews. With four teams per condition, the study was designed to identify qualitative patterns rather than statistical or causal effects. All patterns reported below should be read as qualitative tendencies (see Section~\ref{sec:limit} for further discussion).
%Next, to examine how the timing of AIS use might shape students' practices and perceptions, groups were assigned to conduct interviews in different sequences (indicated by branching arrows). Four groups interviewed their AIS first, then interviewed a human stakeholder (AIS-interview-first sequence, shown in blue). Four groups interviewed a human stakeholder first, then their AIS (human-interview-first sequence, shown in orange). 
%Sequence condition (AIS-interview-first vs. human-interview-first) was assigned to each of the eight groups using a random-number generator, independent of the group's chosen project domain or team composition. Because groups selected their own project topics before sequence assignment, domain was not matched or balanced across the two conditions, and remains a possible alternative explanation for some between-condition differences (as further detailed in the limitations, Section~\ref{sec:limit}). 
%Further, because sequence was assigned at the team level and each team experienced only one sequence, this is effectively a between-groups design with four teams per condition (N=4 per arm). The study was not intended to support statistical or causal claims about sequence effects; all patterns reported below should be read as qualitative tendencies (see Section~\ref{sec:limit} for further discussion).

No group was required to adapt its interview protocol or AIS between interviews. AIS-interview-first teams could use their AIS interview when revising the subsequent human-stakeholder protocol, while human-interview-first teams could use the human interview when revising the subsequent AIS or its protocol. Whether and how teams did so was left to their discretion. These adaptations are therefore treated as observed or reported practices, not as fixed stages of either condition.

This structure allowed us to observe patterns across several dimensions:
\begin{itemize}
    \item How students characterized and differentiated between insights from AIS versus human stakeholders
    \item How interview sequence relates to protocol adaptation, expectations, and sensemaking practices
    \item How insights from each source were reflected in the final design artifacts
    \item How students developed empathy and navigated the cognitive demands of working with synthetic versus authentic voices
\end{itemize}

\subsection{Participants and Ethics}
Twenty-one students (8 groups) participated as part of their required coursework in a semester-long sociotechnical design course. Students provided written informed consent for participation and for the research use of their coursework materials. Data collection for research purposes was voluntary and did not affect course grades. 
Participants were 19 undergraduate and 2 graduate students enrolled in the course, working in 8 teams of two to three students. All students followed the same preparatory sequence before sequence assignment, including course readings on persona construction and evaluation, two short in-class AIS-creation exercises, a guest lecture, and the peer-review evaluation described in Section~\ref{sec:guide}. These common activities provided a shared instructional context, but they do not establish equivalent prior experience, participant experience, or AIS quality across teams. A pre-course measure of prior experience with design methods, generative AI tools, personas, or stakeholder interviewing was not administered. 

After final grades were submitted, students were invited to participate in voluntary post-course qualitative sessions. These comprised one two-participant discussion and one individual interview, both facilitated by a researcher who had not been involved in course instruction.

The study was reviewed and approved by the Institutional Review Board at the authors'  university.

\subsection{Projects Description}
 In the opening two weeks, students introduced themselves through a structured activity, after which they formed groups of two to three members. Here, we outline the projects each group pursued.
\begin{table}[!ht]
\centering
\caption{Student projects by interview sequence, domain, and focus.}
\begin{tabular}{|p{3cm}|p{3.5cm}|p{4cm}|p{4cm}|}
\hline
\textbf{Group} & \textbf{Domain} & \textbf{Interviews} & \textbf{Focus} \\
\hline
\multicolumn{4}{|c|}{{\color{myblue}\textbf{AIS-interview-first sequence}}} \\
\hline
\textit{A1: Mental Health Crisis Support (CareTech)} & Volunteer support systems for crisis hotlines & 
\textit{AIS}: Mental health first-responder \newline 
\textit{Human}: Crisis hotline volunteer & 
Developing technological tools to support crisis hotline volunteers during their active work \\
\hline
\textit{A2: Educational Communication Systems (Teacher System)} & Teacher-parent communication and student management & 
\textit{AIS}: Educator with multiple roles \newline 
\textit{Human}: Special education teacher & 
Reduce administrative burden and improve parent-teacher communication \\
\hline
\textit{A3: Food Delivery Privacy} & Privacy in food delivery platforms & 
\textit{AIS}: Professional operations manager \newline 
\textit{Human}: Company local manager & 
Privacy protection in gig economy applications \\
\hline
\textit{A4: Medical Training Conversational AI} & Medical education and training & 
\textit{AIS}: Medical training participant \newline 
\textit{Human}: Medical professional trainee & 
Conversational AI for medical skill development \\
\hline
\multicolumn{4}{|c|}{{\color{myblue}\textbf{Human-interview-first sequence}}} \\
\hline
\textit{B1: Elderly Transportation Technology} & Public transportation app usability for elderly & 
\textit{Human}: Elderly public transportation user \newline 
\textit{AIS}: Senior citizen using navigation technology & 
Enhance public transportation application access for elderly users  \\
\hline
\textit{B2: Academic Accommodation System (ADHD)} & University support for learning differences & 
\textit{Human}: Student with ADHD \newline 
\textit{AIS}: Student with learning disabilities seeking accommodations & 
Streamlining access to disability services and academic rights \\
\hline
\textit{B3: University Registration System} & Academic course registration & 
\textit{Human}: Third-year science student \newline 
\textit{AIS}: Frustrated student seeking interdisciplinary courses & 
System barriers to cross-disciplinary learning \\
\hline
\textit{B4: Gaming Toxicity Management} & Online gaming community behavior & 
\textit{Human}: Experienced gamer \newline 
\textit{AIS}: ``AggroMaster'' – competitive gamer prone to toxic behavior & 
 A general reward-based system to encourage non-toxic behavior and teamwork in gaming communities \\
\hline
\end{tabular}
\label{tbl:projects}
\end{table}

Table~\ref{tbl:projects} details the projects selected by the student groups, their focal goals, and the order in which groups engaged with AIS and human stakeholders. Project selection began with an initial literature review, during which students surveyed existing solutions and articulated the societal and stakeholder needs motivating their chosen focus. Over the semester, groups documented their progress through staged submissions aligned with the course structure. A final report synthesizing their design process and proposed intervention was submitted at the end of the semester.
\subsection{Data Collection and Analysis}
\label{subsec:datacollection}

Our analysis drew on two complementary approaches: a structured coding procedure applied to course-embedded materials to identify correspondences between interview-derived concerns and final design elements, and thematic analysis of one post-course two-participant discussion, one post-course individual interview, and the end-of-course class discussion. %The correspondence analysis does not establish that an interview statement caused a design decision.

%\noindent \paragraph{\textbf{Course-embedded data coding.} } 

Throughout the semester, student groups submitted staged deliverables, including interview transcripts, design artifacts, and reflection documents.  These materials were analyzed through multiple coding procedures.

 The students team is the primary unit of analysis, and later quotations in the paper are labeled by group (A1–A4, B1–B4). Where an individual speaker within a group is quoted directly, the identifier refers to a maintained participant internal documentation. 

%\noindent \textit{Pain point identification and classification.} To identify the pain points discussed in the human and AIS interviews for each of the groups, one of the authors, i.e., primary coder,  independently conducted a systematic review of each group's AIS and human stakeholder interview transcripts alongside their final design submissions. The researcher first identified pain points in the transcript using the thematic analysis approach of Braun and Clarke \cite{braun2013successful}. The process began with a familiarization phase in which the researcher read all final reports and design artifacts to develop initial observations about each group's conceptual space before engaging with the interview transcripts.
%Following familiarization, the researcher identified pain points raised in each interview, including both AIS and human stakeholders and  developed initial codes to capture their perceived prominence. In this phase, a pain point was considered \textit{prominent} when it appeared repeatedly across responses or was expressed with clear intensity and treated as a central theme in the interview. It was considered \textit{mentioned} when it appeared briefly or in passing without further elaboration. This distinction reflects how salient each issue was within the interview itself. In addition, the source of each pain point, whether AIS, human stakeholder, or shared by both, was recorded throughout for each pain point. 
%The coding process was discussed with the other researchers, resulting in fine-tuning the coding done by the primary coder.   
\subsubsection{Pain point identification and classification}
%\noindent \textit{Pain point identification and classification.} 
One of the authors, the primary coder, independently reviewed each group's AIS and human stakeholder interview transcripts alongside their final design submissions, using the thematic analysis approach of Braun and Clarke \cite{braun2013successful}. The process began with a familiarization phase in which the coder read all final reports and design artifacts to develop initial observations about each group's conceptual space before engaging with the transcripts, as the variety of topics and the different approaches were needed to receive enough context. The coder then identified the pain points raised in each interview and coded their perceived prominence. A pain point was considered \textit{prominent} when it appeared repeatedly across responses or was expressed with clear intensity and treated as a central theme; it was considered \textit{mentioned} when it appeared briefly or in passing without elaboration. This distinction reflects how salient each issue was within the interview itself. The source of each pain point, that is, AIS, human stakeholder, or both, was also recorded. The coding was then discussed with the other researchers, resulting in refinements to the primary coder's scheme.

\noindent \paragraph{\textbf{Tracing pain points to design decisions.}} 
%After the initial screening of pain points in the interviews, two researchers identified their influence on the design. 
%\subsubsection{Tracing pain points to design decisions}
After identifying pain points in the interviews, the primary coder compared each pain point with the corresponding team’s reports and final design artifact. Each pain point was assigned to one of four categories that distinguished no identified design correspondence, contextual reference, researcher-inferred feature correspondence, and student-reported foundational influence. The third category records a correspondence identified by the researcher without establishing that the pain point caused the design component. The fourth category requires an explicit statement from the team that the pain point informed or drove the design decision.

\begin{table}[ht]
\centering
\caption{Categories used to assess whether and how interview pain points were reflected in each team’s final design}
\begin{tabular}{p{4.3cm} p{9cm}}
%\hline
\textbf{Category} & \textbf{Criteria for assigning the category} \\
\hline
\textbf{No identified design correspondence} &  No team statement or final design component was identified as addressing the pain point. \\\hline
\textbf{Contextual reference} &  Students referred to the pain point as context or background, but no specific design component addressed it. \\\hline
%\textit{Noticeable Influence}& Students referred to the pain point  in their documents and the design included features that noticeably reduce the difficulty or friction associated with the pain point. While the root problem may still exist, the user's process for navigating it is made tangibly easier. \\\hline
\textbf{Researcher-inferred feature correspondence} &  The final design contained a specific component that addressed the pain point, but the team did not explicitly state that the pain point prompted the component. This category records correspondence, not causal influence. \\\hline
\textbf{Student-reported foundational influence} &  A specific design component addressed the pain point, and the team explicitly stated that the pain point informed or drove that design decision. \\\hline
\end{tabular}
\label{tbl:pp-criteria}
\end{table}
Table~\ref{tbl:pp-criteria} details the categories. 
Cases coded as ``No identified design correspondence'' or ``Contextual reference'' were not assigned an intervention level. Cases coded as either ``Researcher-inferred feature correspondence'' or ``Student-reported foundational influence'' were then classified by the level of the corresponding design intervention. Intervention level describes the scale of the design element, not the strength or causal magnitude of the interview’s influence.
%First, influence was coded on a four-level scale: 1 -- no influence,  2 -- indirect influence, 3 -- direct influence and 4 --stated foundational influence. The categories captured whether a pain point was noticed but merely acknowledged, informed the overall direction, or directly generated a specific design component. 
%Researchers compared their notes and discussed and reached agreement in the few cases of disagreement.

\noindent \paragraph{\textbf{Design Interventions.}} 
The process of identifying the levels of design interventions followed similar steps. Similar to the pain point coding process, this part was also discussed among the other researchers. Table~\ref{tbl:pp-criteria1} details the coding criteria for the design interventions level. 

\begin{table}[ht]
\centering
\caption{Coding criteria for levels of design intervention}
\begin{tabular}{p{3cm} p{10cm}}
%\hline
\textbf{Level} & \textbf{Design Intervention} \\
\hline
1 & Interface adjustments (e.g., button placement, text size). \\
2 & Feature modifications (e.g., new functionality, workflow changes). \\
3 & Conceptual redesign (fundamental approach alterations). \\
%\hline
\end{tabular}
\label{tbl:pp-criteria1}
\end{table}

\noindent \paragraph{\textbf{Intercoder Reliability.}} 
In addition to the research team discussions of the pain-point coding and design-correspondence coding, we assessed the reliability of the coding scheme on a subsample of the data. We trained a research assistant (referred to as a secondary coder) on the coding scheme through a briefing on the study's aims and coding categories, followed by a calibration exercise in which they reviewed the coding of one project to familiarize themselves with its application. Following this training, they independently coded pain-point prominence and the design-correspondence/attribution category for four of the eight groups.  We restricted the reliability comparison to pain points that both coders identified. Cases where only one of the coders identified a pain point beyond those raised by the students themselves were excluded from this comparison. 
\begin{table}[h]
\centering
\caption{Intercoder reliability between the original coder and an
independent research assistant on four of the eight groups}
\label{tab:reliability}
\begin{tabular}{llc}
\toprule
Category & Measure & Krippendorff's $\alpha$ [95\% CI] \\
\midrule
Pain Points & Prominence & 0.746 [0.488, 0.942] \\
Pain Points & Design-correspondence/attribution category & 0.850 [0.705, 0.955] \\
Design & Intervention Level & 0.849 [0.571, 1.000] \\
\bottomrule
\end{tabular}
\end{table}
Table~\ref{tab:reliability} reports the intercoder-reliability results for pain-point prominence, design-correspondence/attribution category, and intervention level.
Agreement was substantial for the two pain-point measures: Krippendorff's $\alpha = 0.746$ [0.488, 0.942] for prominence and $\alpha = 0.850$ [0.705, 0.955] for the design-correspondence/attribution category. Agreement for intervention level was $\alpha = 0.849$ [0.571, 1.000]. These coefficients assess agreement in prominence, design correspondence/attribution, and intervention level conditional on both coders having identified the pain point.

%illustrates the analytic process in the coding of group B1's project, which redesigned a public transportation application for elderly users. 

%\hline

The full code book for group A1 appears in Appendix~\ref{sec:apdxA}.  The analytic process in the coding of group B1's project, which redesigned a public transportation application for elderly users, is illustrated in Appendix~\ref{ref:apdx} .
\begin{comment}
\begin{itemize}  
    \item An independent research assistant analyzed how insights derived from human and AIS interactions were reflected in each group’s final design outcomes by identifying pain points, classifying their prominence, and tracing how they were taken up in design decisions. Pain points were classified as \textbf{\textit{prominent}} if they were central in an answer, and recurred more than once, and as \textbf{\textit{mentioned}} if they were referenced within a few answers without being the primary focus. The research assistant coded all interviews and final works.
    \item AIS creation reports, that included students' reflections, were evaluated by a researcher not involved in course instruction.  
    \item Researchers read and discussed all students reports.
\end{itemize}  
This enabled systematic examination of how insights from each interview type were reflected in design decisions.
\end{comment}

\subsubsection{Written reflections}  We analyzed 13 individual written reflections submitted by 13 of the 21 students, representing seven of the eight project teams. Three students also participated in the post-course sessions. Across the reflection and post-course sources, all eight project teams were represented, although coverage was uneven. %To reduce the risk of deductive identification in this small cohort, we do not link voluntary post-course participation to specific project teams.
\subsubsection{Post-course discussion, individual interview, and class discussion}  
Two post-course meetings were conducted in Hebrew with three participants approximately four months after the course. The two-participant discussion included students from teams A2 and A3, and the individual interview included a student from team A1. All three teams were in the AIS-interview-first condition. The meetings were conducted over Zoom, lasted approximately 32 minutes on average, and represented a 14.3\% participation rate. They were audio-recorded, automatically transcribed using the Hebrew AI platform, and checked against the recordings during analysis. Because the post-course sample contained no human-interview-first participant, these sessions were used to deepen accounts of AIS construction, evaluation, trust, and preparatory use within the AIS-interview-first condition; they were not used to establish differences between the two conditions.
Cross-condition comparisons were conducted separately at the project-team level using course-embedded materials and the end-of-course class discussion, which was reorganized by project team. The post-course sessions therefore supplied elaboration and qualifying cases rather than representative coverage of both conditions. Quotations are labeled using pseudonymous participant identifiers (e.g., P2 and P4). To protect participant confidentiality, quotations are identified only through pseudonymous participant and team codes. We nevertheless distinguish whether evidence came from the two-participant post-course discussion, the individual post-course interview, the end-of-course class discussion, or a written reflection.

To analyze these materials, one researcher first analyzed the two-participant post-course discussion data, taking a hybrid deductive–inductive approach~\cite{fereday2006demonstrating}. An initial set of codes derived from the shared post-course discussion and interview protocol was expanded inductively and consolidated into the codebook. The resulting codebook was then applied to the individual post-course interview, the end-of-course class discussion, and the 13 written reflections. Appendix~\ref{sec:interviewprotocol} provides the protocol, and Appendix~\ref{app:qualitative-codebook} presents the resulting codebook.    

%Specifically, RQ1 findings draw on students’ individual written reflections and the end-of-course class discussion, supplemented by two voluntary post-course sessions conducted approximately four months later. The post-course material comprised one two-participant discussion and one individual interview. Because linking individual excerpts to either session would increase the risk of deductive identification in this small cohort, particularly because participant codes are associated with project teams, we report the composition and timing of the sources in aggregate but do not identify the specific session from which each excerpt was drawn.

%The scaling and rating criteria for the influence of pain points on the design components and the design component levels are detailed in  Table~\ref{tbl:pp-criteria}. 

\section{AIS Creation Process}
\label{sec:ais}
Before examining patterns in how interview sequence related to design outcomes, we first describe the AIS creation process itself. 

\subsection{Guidelines}
\label{sec:guide}
Students engaged with core readings on persona construction and evaluation, which were discussed in class to establish a shared vocabulary and quality standards~\cite{salminen2021survey,goel2023preparing}. To situate the exercise in current practices, we hosted a guest lecture by the CEO of a startup that deploys a human resources AIS that supports managers in matching tasks to employees and in structuring team work. In addition, before starting the work on the course project, students tried two brief, informal in-class AIS-creation exercises to gain hands-on familiarity with the tools and vocabulary. These were exploratory rather than formally structured.

\paragraph{Developing an AIS Evaluation Framework. }
We established a standardized set of AIS-evaluation criteria through structured class discussions informed by the literature~\cite{salminen2021survey,goel2023preparing}. All student groups applied the same five dimensions: (1) \textbf{credibility:} believability and evidential grounding of user characteristics; (2) \textbf{consistency:} internal coherence across persona elements; (3) \textbf{completeness:} coverage of salient population attributes and identification of gaps, with sufficient detail for design decisions; (4) \textbf{clarity:} well-organized, comprehensible presentation; and (5) \textbf{empathy/depth:} motivations and constraints that foster user understanding.
Applying the same evaluation dimensions gave teams a shared evaluative vocabulary. It did not establish equivalent AIS quality, prompting practices, prior experience, or participant experience across teams. Differences observed later in the course are therefore treated as exploratory patterns associated with the complete sequencing conditions as implemented.

\subsection{Initial creation process of the AIS}
Within this context, students proceeded to generate the AISs.  Groups began by defining basic demographic and contextual parameters for their target users. The Teacher Management System group, for instance, started with specifications like ``30-year-old female, mother of one, married for 5 years, 4 years of school experience, homeroom teacher for special education class, with additional responsibilities as science coordinator''. 
Several groups documented a struggle with their initial efforts of crafting prompts, with one group asking the LLM directly to give them advice on the prompt. Character consistency emerged as another problem. The Elderly Transportation group performed a simple test, asking their created AIS for its name. In response, the AIS answered with a new name that differed from the originally established one.

Once the students felt their AIS met the set criteria defined, the class entered a peer-review step, where groups evaluated other groups' AISs, as follows. 

\subsection{Peer Review Evaluation }
To evaluate the created AISs, students employed a peer-review interactive evaluation process 
using De Bono's Six Thinking Hats methodology~\cite{debono1985six}, with groups rotating through different 
evaluative perspectives and challenging each other's creations. This peer review occurred during the joint preparation phase, before assignment to the two interview-order conditions. It provided a common review activity but did not ensure that the resulting AIS were equivalent in quality, specificity, refinement, or students’ experience of working with them. This structured peer review process revealed 
both the potential and limitations of AISs in ways that individual assessment might have missed.

Students iteratively evaluated their AISs using multiple peer-review lenses, which surfaced recurring limitations of synthetic stakeholder representations. Fact-oriented review highlighted gaps in evidentiary grounding, prompting teams to distinguish assumption-based from evidence-based AIS characteristics and to supplement synthetic responses with quantitative data and real-world examples. Emotion-focused review revealed tensions between surface plausibility and emotional authenticity, leading students to adjust how AISs articulated motivations, frustrations, and values. Critical review proved particularly consequential, exposing patterns of overgeneralization and stereotyping; for example, teams were prompted to recognize diversity within stakeholder groups and to reconsider simplified or implausible narratives embedded in their AISs. Together, these evaluations foregrounded how synthetic stakeholder voices tend to abstract, generalize, or smooth over lived complexity, requiring deliberate interrogation to avoid misrepresenting stakeholder perspectives.

\section{Results}
Appropriation is examined across several parts of the study rather than as a single outcome. Section 4 documents how students constructed and evaluated AIS. Section~\ref{sec:rq1} addresses the interpretation and evaluation components of RQ1. Section 5.2 addresses RQ3 by identifying correspondence between interview-derived concerns and final design elements while distinguishing researcher inference from explicit student attribution. Section 5.3 primarily addresses RQ2 through reported adaptation, validation, challenge, and use across the two interview-order conditions. Together, these sections address the construction, adaptation, interpretation, evaluation, integration, and rejection practices included in our definition of appropriation.

%The Results are organized by each subsection's primary research-question alignment. Section~\ref{sec:rq1} addresses RQ1 by examining how students interpreted, evaluated, and used AIS input. Section~\ref{sec:rq3} addresses RQ3 by tracing interview-derived concerns into final design decisions. Lastly, Section~\ref{sec:rq2} addresses RQ2 by comparing the reported practices and interpretations of AIS-interview-first and human-interview-first teams. Because appropriation, sequencing, and design carry-through are related, some findings inform more than one RQ, but each is presented under its primary analytic role.

\subsection{Differences in How AIS and Human Stakeholder Input Was Interpreted}
\label{sec:rq1}
%We address here RQ1 using the eight teams' written reflections, supplemented by three participants' post-course focus-group accounts. Teams are the primary comparative unit, individual comments provide supporting evidence, and source-level patterns were identified through the qualitative comparison described in Section~\ref{subsec:datacollection}.
This section addresses the interpretation and evaluation components of RQ1 using individual reflections, the end-of-course class discussion, and post-course interviews. Because the post-course participants were all from AIS-interview-first teams, their accounts are used only to elaborate that condition.

When students compared AIS interviews with interviews conducted with human stakeholders, they consistently described differences in how each source could be interpreted and used. These differences appeared across the qualitative corpus, including the written reflections, end-of-course class discussion, two-participant post-course discussion, and individual post-course interview, and concerned the level of specificity, the focus of the response, the expression of uncertainty and emotion, and the extent to which responses remained close to lived experience.

\textit{General framing.}
Students often described AIS as echoing common patterns while lacking the specificity, context, and nuance they encountered in human stakeholder interviews. Half of the groups referred to AIS responses as general, and sometimes stereotypical. As P16 from Group B3 put it in their reflection: ``\textit{It was very vanilla, not entirely accurate}.''

For two groups (B4, A2), this generality was closely tied to a lack of contextual background. Group B4, which examined gaming toxicity, noted that the AIS described toxic behavior in online games in broad terms, but did not distinguish between the different settings in which such behavior appears. The human stakeholder, by contrast, described these differences in a more situated way, explaining that non-competitive games were largely free of toxic behavior, while competitive environments could become overwhelming. The participant even referred to one specific game and said, ``\textit{I couldn't handle the toxicity there, they are especially toxic to newcomers}.''

Students also described AIS as producing the kinds of answers they expected in advance, rather than surfacing the unusual or revealing details they were hoping to uncover. Group B3 (P16) explained:
\begin{quote}
``\textit{It was very stereotypical, giving us the kind of answer it assumed a typical person would say. But it didn’t behave like a real person overall. What interested us, though, were not the stereotypical things, but the edges, the differences, the pain points.}''
\end{quote}

Relatedly, some students suggested that AIS did not have enough background to provide contextually adequate responses. Referring to existing solutions in the CareTech domain, one group noted that the AIS appeared to respond from partial knowledge rather than from a situated understanding of the field.

Taken together, these accounts suggest that AIS often provided a broad initial framing of the problem space, but one that students found difficult to prioritize or trust without the contextual grounding that emerged in human interviews.

\textit{Focus and Depth.}
Three groups (B1, B2, B3) discussed AIS's difficulty in aligning its focus with what mattered most. Two groups (B1, B2) explained that AIS sometimes emphasized peripheral issues while overlooking human stakeholders' central concerns, or, as described by group B3, the AIS ``decided'' to focus on one aspect only. P16 from B3 described: ``It took one aspect from the whole description of the cases and the problem, and focused only on that, without addressing anything else''. Continuing the stereotypical perspective mentioned earlier, here, Group B2 deepens their discussion on how AIS focuses entirely on specific aspects that are commonly discussed in the context of people with ADHD. 
In their reflection, P13 (B2) contrasted this with human interviews, where lived experience often made the central problem more concrete and actionable.

\begin{quote}``\textit{It saw the word ADHD and focused on how people with ADHD need help with concentration and maybe time management. It started talking about those things, and completely ignored the real struggles the [human] student described}''. 
\end{quote}

\noindent For example, the stakeholder described great difficulties in communicating with the system and going through the process: 

\begin{quote} ``\textit{The process took me a year and a half and I had to chase after them, it's hard to reach them and I didn't know at all who I needed to talk to at first, and each time they referred me to someone else}''.   
\end{quote}

\noindent While the AIS stated a preference: 

\begin{quote} ``\textit{I would also be happy if there was an option to do the process in a more smooth, flowing and fast way}''.  
\end{quote}
%selection of significant topics, and how this selection sometimes is not aligned with the human stakeholders' main pain points

\textit{Certainty.}
A smaller but important set of groups criticized AIS for sounding overly certain. Two groups (A1, A3) described the AIS as polished, structured, and confident, even when uncertainty would have been more appropriate. During the end-of-course class discussion, Group A1 similarly described: ``\textit{The AIS always gave neat bullet-point answers, but real people spoke with hesitation and doubt}.''

In the food delivery project (A3), students felt that the AIS responded in an evasive, public-relations style that protected the organization rather than exposing the underlying privacy problem. P6 described it this way, as part of the personal reflection:
\begin{quote}
``\textit{It was very political, giving answers in line of [describing AIS ``internal perspective'']: `I don’t want to damage the company’s image, so I won’t say anything. In fact, I’ll say positive things instead'. But we were aiming to solve privacy problems in this company, and it responded: `Such problems do not exist, and everything is fine.'}''
\end{quote}

By contrast, students described human stakeholders as more likely to hesitate, qualify their statements, or contradict themselves. These qualities were not treated as weaknesses. Rather, students often interpreted them as signs that the interview reflected lived experience and as markers of authenticity.

\textit{Emotional intensity.}
Across all groups, the human interviews carried a stronger emotional impact, as stakeholders conveyed their pain points with evident emotional intensity. They employed explicitly emotional terms and, at times, repetitive language or fragmented speech, features that may signal stress. Stakeholders also openly shared their feelings. For instance, the CareTech participant (Group A1) emphasized that user feedback was crucial for her self-esteem and motivation to remain in the job; the participant with ADHD (Group B2) described the persistent emotional difficulty of interacting with the system; and the gamer (Group B4) reflected on his feelings during gameplay:

\begin{quote}
``\textit{...ruin motivation. It's a buzz-kill... makes me regret playing}''.
\end{quote}

The Educational communication System participant (P4, A2) noted during the two-participant post-course discussion: 
\begin{quote}
    ``\textit{The emotional aspect, which was missing in the AI-generated responses, became evident in the real interview, with the participant revealing frustration, exhaustion}''.
\end{quote}
For students, this emotional texture made human interviews harder to simplify, but also more memorable and more consequential for later design decisions.

 \textit{Solution-oriented.}
While primarily aiming to learn about pain points through the interviews, students noted that AISs did not merely recount pain points, but they reformulated them as problem–solution pairs. This sometimes resonated with students but in other cases misrepresented the raw experience. As noted by Group B2, the ADHD project, the AIS reframed frustration as a neat fix:``The AIS reframed the complaint into: `If the rights were explained clearly on the website, it would be much easier'.''  Yet, human stakeholders expressed their confusion and exasperation more directly, which students found more relatable and easier to follow. The solution-oriented approach was also a possible source of delay, as in cases where the AIS’s suggested focus later proved to be misguided, as reflected  by P6 (Group A3): ``Unfortunately, in our case, it didn’t contribute to the solution at all and actually delayed us.''

Taken together, these accounts suggest that AIS often helped students begin to frame a problem space, but did so in a smoother and more generalized way. Human interviews were narrower in scope, yet more grounded, more emotionally textured, and more likely to redirect students toward concerns they had not prioritized in advance.

\subsection{Design Outcomes} 
\label{sec:rq3}
This section addresses RQ3 using all eight teams' AIS and human-stakeholder interview transcripts and final design artifacts. Each interview-derived pain point is the coding unit within a team, and pain points were coded for source and prominence and then compared with the final designs to identify design correspondence, explicit student attribution, and - where a specific design correspondence was identified - the intervention level using the structured procedure in Section~\ref{subsec:datacollection}.

\begin{table}[htbp]
\small
\setlength{\tabcolsep}{4pt}
\centering
\caption{Pain points by source and prominence, and researcher-coded correspondence with final design elements.}
\label{tbl:designpoints}
\begin{tabularx}{\linewidth}{l*{5}{C}*{2}{C}}
\toprule
\textbf{Group} &
\multicolumn{5}{c}{\textbf{Interview Pain Points}} &
\multicolumn{2}{c}{\textbf{Final-Design }} \\
\cmidrule(lr){2-6}\cmidrule(lr){7-8}
 & \multicolumn{2}{c}{\textbf{AIS}} 
 & \multicolumn{2}{c}{\textbf{Human}} 
 & \multirow{2}{=}{\centering\textbf{Prominent Shared}}
 & \textbf{AIS} 
 & \textbf{Human} \\
\cmidrule(lr){2-3}\cmidrule(lr){4-5}
 & \textbf{mentioned}
 & \textbf{prominent}
 & \textbf{mentioned}
 & \textbf{prominent}
 & 
 & \textbf{$n/N$ (\%)}
 & \textbf{$n/N$ (\%)} \\
\midrule
A1.\ CareTech
  & 2 & 2 & 4 & 7 & 0
  & 0/2 (0.0\%)
  & 5/7 (71.4\%) \\

A2.\ Teacher System
  & 1 & 7 & 2 & 7 & 2
  & 4/7 (57.1\%)
  & 6/7 (85.7\%) \\

A3.\ Food Delivery
  & 1 & 1 & 2 & 5 & 0
  & 1/1 (100\%)
  & 4/5 (80.0\%) \\

A4.\ Medical Training
  & 0 & 4 & 0 & 3 & 0
  & 4/4 (100\%)
  & 3/3 (100\%) \\
\midrule
B1.\ Elderly Transportation
  & 0 & 5 & 0 & 4 & 4
  & 5/5 (100\%)
  & 4/4 (100\%) \\

B2.\ Academic Accommodation
  & 3 & 1 & 2 & 3 & 1
  & 1/1 (100\%)
  & 3/3 (100\%) \\

B3.\ University Registration
  & 4 & 5 & 3 & 6 & 1
  & 3/5 (60.0\%)
  & 6/6 (100\%) \\

B4.\ Gaming Toxicity
  & 11 & 7 & 4 & 8 & 4
  & 4/7 (57.1\%)
  & 8/8 (100\%) \\
\midrule
\textbf{Overall}
  & \textbf{22}
  & \textbf{32}
  & \textbf{17}
  & \textbf{43}
  & \textbf{12}
  & \textbf{22/32 (68.8\%)}
  & \textbf{39/43 (90.7\%)} \\
\bottomrule
\end{tabularx}

\vspace{0.35em}
\parbox{\linewidth}{%
\footnotesize
\textit{Note.} Final-design correspondence is reported as $n/N$,
where $n$ is the number of prominent pain points for which a
corresponding final-design element was identified and $N$ is the total
number of prominent pain points produced by that source. Shared pain points are included in both source-specific totals. The percentages therefore describe overlapping source-specific correspondence rates, not independent statistical comparisons.%The percentages describe source-specific carry-through; they do not measure causal influence, decision weight, or the magnitude of the corresponding design change.
}
\end{table}

\begin{comment}
\begin{table}[htbp]
\small
\setlength{\tabcolsep}{4pt}
\centering
\caption{Pain points by source and prominence, and researcher-coded correspondence with final design elements.}
\label{tbl:painpoints}
\begin{tabularx}{\linewidth}{l*{5}{C}*{2}{C}}
%\toprule
\textbf{Group} &
\multicolumn{5}{c}{\textbf{Interview Pain Points}} &
\multicolumn{2}{c}{\textbf{Final Design}} \\
\cmidrule(lr){2-6}\cmidrule(lr){7-8}
 & \multicolumn{2}{c}{\textbf{AI}} & \multicolumn{2}{c}{\textbf{Human}} & \multirow{2}{=}{\textbf{Prominent Shared}} & \textbf{AI} & \textbf{Human} \\
\cmidrule(lr){2-3}\cmidrule(lr){4-5}
 & \textbf{mentioned} & \textbf{prominent} & \textbf{mentioned} & \textbf{prominent} & & \textbf{prominent} & \textbf{prominent} \\
\midrule
A1.\ CareTech                  & 2 & 2 & 4 & 7 & 0  & 0 & 5 \\
A2.\ Teacher System            & 1 & 7 & 2 & 7 & 2 & 4 & 6 \\
A3.\ Food Delivery             & 1 & 1 & 2 & 5 & 0 & 1 & 4 \\
A4.\ Medical Training          & 0 & 4 & 0 & 3 & 0 & 4 & 3 \\
\midrule
B1.\ Elderly Transportation    & 0 & 5 & 0 & 4 & 4 & 5 & 4 \\
B2.\ Academic Accommodation    & 3 & 1 & 2 & 3 & 1 & 1 & 3 \\
B3.\ University Registration   & 4 & 5 & 3 & 6 & 1 & 3 & 6 \\
B4.\ Gaming Toxicity           & 11 & 7 & 4 & 8 & 4 & 4 & 8 \\
\bottomrule
\end{tabularx}
%\vspace{.5em}
\label{tbl:designpoints}
\end{table}
\end{comment}

Table 5 reports both the number of prominent pain points identified from each source and the number for which a corresponding element was identified in the final design, describing source yield and carry-through. Each pain point is counted as one unit regardless of whether the corresponding design element was an interface adjustment, feature modification, or conceptual redesign. We henceforth refer to such design elements as corresponding to a pain point. 

We further see that final-design correspondences were identified for 22 of 32 prominent AIS-derived pain points (68.8\%) and 39 of 43 prominent human-stakeholder-derived pain points (90.7\%). 
In the AIS-interview-first condition, the number of human-derived correspondences was higher in three of the four teams: A1 had 0/2 AIS-derived and 5/7 human-derived prominent pain points with design correspondence; A2 had 4/7 and 6/7; and A3 had 1/1 and 4/5. In A4, all prominent pain points from both sources had design correspondence—4/4 AIS-derived and 3/3 human-derived—even though the raw counts differed. Thus, the condition does not show a uniform source advantage when incorporation is considered relative to the number of prominent concerns each source produced.
In the human-interview-first condition, both sources had complete incorporation in B1 (5/5 AIS; 4/4 human) and B2 (1/1 AIS; 3/3 human). B3 had correspondence for 3/5 AIS-derived and 6/6 human-derived prominent pain points, while B4 had 4/7 and 8/8, respectively. These patterns indicate differences in the frequency with which coded concerns corresponded to final design elements; they do not establish how much decision weight students assigned to each source or whether one source caused a larger design change.

Students who started with stakeholders reflected that beginning with authentic voices helped them refine their use of AISs (P19, group B4):  
\begin{quote}
``\textit{The interview [with the human] helped us sharpen our understanding of the main difficulties … collecting and organizing the information gathered in the interview for designing the AIS so that it fully reflects the experiences, needs and challenges.}''
\end{quote}

Taken together, these patterns suggest different roles for AIS across the two sequences. In AIS-interview-first teams, AIS often provided an early frame that was later narrowed, revised, or displaced by human input. In human-interview-first teams, AIS was more often used to revisit, validate, or extend issues that had already been grounded in stakeholder experience. These are best understood as qualitative tendencies in how students worked with different sources, not as claims about fixed effects.

\subsection{Sequencing Patterns in Stakeholder Engagement}
\label{sec:rq2}
This section primarily addresses RQ2 by comparing reported practices and interpretations across the four AIS-interview-first and four human-interview-first teams. The cross-condition comparison draws on the verified team-level course materials and end-of-course class discussion. The three post-course participants (all from AIS-interview-first teams) are used only to elaborate accounts within that condition. The team is the comparative unit, and the findings are exploratory patterns associated with the conditions as implemented.
%This section primarily addresses RQ2 by comparing the four AIS-interview-first teams with the four human-interview-first teams using team reflections, supplemented by post-course focus-group accounts. The team is the comparative unit, and reported interpretations and interview-practice adaptations were examined through the cross-sequence qualitative comparison described in Section~\ref{subsec:datacollection}.

Students’ accounts revealed recurring differences in how AIS and human stakeholder interviews were interpreted and used, depending on the order in which they occurred. Students’ accounts differed across the conditions in their interview practices and in how they described the contribution of each source.%sequence was associated not only with different interview practices, but also with different ways of assigning value to what each source contributed.

Groups in the AIS-interview-first sequence (A1–A4) often found the AIS's answers wrong or overly general, such as focusing on mental health difficulties that the human stakeholder did not confirm (group A1), or claiming there were no privacy problems in the company, which the stakeholder utterly contradicted (group A3). Group A2 summarized it as follows: 
\begin{quote}
    ``\textit{While the AIS captured broad themes, it lacked nuance and prioritization. The simulated responses tended to generalize issues rather than highlight the most pressing concerns, making it difficult to determine which pain points were most impactful.}''
\end{quote}
Still, the groups emphasized the preparatory value of the AIS: it showed whether their questions could elicit substantive answers, and sometimes introduced new perspectives. All AIS-interview-first groups described the AIS as a ``good preliminary check'' that helped them reshape questions. One student (group A4) explained that ``it included several points we hadn’t thought of ourselves''.  During the end-of-course class discussion, a student from B4 explained that using AIS creation in another course gave them a mental model of a store owner and ``broke the fear'' of the first interview. During the same class discussion, students from A2 and B3 described extending this use to job-interview preparation and simulations of workplace conversations. In these cases, AIS did not replace stakeholder engagement. Instead, it often served as an initial frame through which students tested questions, rehearsed interviews, and entered the human interview with a more concrete, if still provisional, sense of the domain.

By contrast, groups who began with human interviews (B1–B4) described the AIS as primarily useful for validation. After speaking with human stakeholders, they turned to the AIS to check and repeat themes already observed. For instance, students in group B3 noted that the AIS could ``validate themes that also appeared with the human'', while group B1 reflected that it helped them ``think through possible problems in a systematic way''. In most cases, AISs seldom added novel information, functioning less as tools for discovery and more as predictable mechanisms for confirmation. A notable exception was B1, who reported that by deliberately contrasting stakeholder demographics in the AIS creation, they were able to identify an additional pain point that the human stakeholder had not raised. These observations highlight that the perceived role of AISs in the human-interview-first sequence was generally confirmatory, with occasional exceptions, and should be interpreted as impressionistic patterns rather than quantitative results. In these accounts, AIS rarely functioned as the primary source of discovery. More often, it was used after the human interview as a secondary source that helped students revisit themes, check whether concerns recurred, or think through possible extensions of what they had already heard.

\subsubsection{Patterns in How Students Interpreted and Used Stakeholder Input}

The comparison identified recurring contrasts in how students described processing AIS and human-stakeholder input.
%The analysis of interview order surfaced fundamental differences in how students processed AIS versus human interactions, which was associated with differences in how students described their design reasoning.  

\textbf{Effort and filtering.}
Students in the AIS-interview-first sequence described the AIS interview as requiring substantial effort to make the responses usable. They reported having to narrow prompts, request specificity, and actively filter broad or verbose answers. As one student explained, they ``had to constantly narrow and feed the AI specifics to get anything precise'' (P2, A1). AIS responses were often experienced as covering a wide range of possible issues, but at a level of generality that required additional work to identify what was relevant.

When these students later turned to human interviews, they described a different kind of effort. Human accounts were ``less structured'' and sometimes ``messy'' (A1–A4), yet they were also more emotionally engaging and more clearly organized around a small number of pressing concerns. Rather than filtering excess information, students described the effort as focusing on interpreting emotion, contradiction, and emphasis, and deciding how to carry these forward into their designs.

Students in the human-interview-first sequence described a different progression. Groups that began with a human stakeholder described confronting contradiction, emotion, and contextual detail from the outset. Although challenging, this initial encounter helped them identify what mattered most in the problem space. When they later turned to AIS, they approached it more strategically, often using it to revisit or expand on concerns that had already become salient. In this sequence, Group B3 described the AIS as ``an anchor'' used for validation or comparison rather than as an initial starting point.

Across both sequences, students described AIS as a reference point, but the role of that reference point differed with timing. In AIS-interview-first groups it often set an initial frame that was later revised through human input. In human-interview-first groups it more often supported comparison and validation of insights that were already grounded in stakeholder experience.

\textbf{Trust and credibility.}
Students often described AIS responses as requiring confirmation before they could be taken seriously. Several groups explicitly noted that they did not fully trust AI responses ``until they’re corroborated by a real interview'' (A2, A3). By contrast, human interviews were treated as immediately credible, even when they were incomplete, hesitant, or internally inconsistent.

Students also reported that inconsistencies in AIS responses were harder to detect. Group A1, for example, described subtle contradictions in the AIS replies that were not immediately visible but became apparent only after closer inspection. This further reinforced a cautious stance toward AIS, where responses were treated as provisional unless they aligned with stakeholder accounts.

These accounts also differed in how students reported using each source in their design process. Insights from human stakeholders were more readily carried forward into design decisions, while AIS responses were more often revisited, checked, or reformulated. 
Some human-interview-first teams reported adjusting subsequent AIS prompts or interview strategies to reflect concerns raised by the human stakeholder. Because refinement was discretionary, this is treated as a reported practice rather than a fixed feature of the condition.

\textbf{Challenging assumptions.}
Human interviews frequently surfaced issues that students had not prioritized in advance, including contradictions and non-obvious concerns that unsettled their assumptions. As several groups noted, these interviews revealed ``non-obvious and sometimes contradictory insights'' that forced them to reconsider their initial framing.

AIS, by contrast, was often described as returning expected or socially recognizable answers. Students across multiple groups noted that the AIS tended to ``tell us what we wanted to hear'' (A1–A4, B2–B4), reinforcing rather than challenging their existing assumptions.

In this sense, AIS more often supported an initial frame, whereas human interviews were more likely to redirect it by introducing unexpected priorities or tensions.

\textbf{Emotional texture.}
Students repeatedly described human interviews as carrying what one group called ``messy truths and emotion'' (A3). Human stakeholders expressed frustration, exhaustion, hesitation, and repetition, and these qualities made the interviews harder to simplify into neat summaries. AIS responses, by contrast, were often described as ``polite, non-committal'' (A2, A3, B3), and as producing accounts that felt ``not grounded in reality'' (A1–A4, B2, B3, B4).

This difference was reflected in how specific concerns were articulated. In the Medical Training project (A4), the AIS discussed general ethical considerations around medical data. The human stakeholder, however, repeatedly emphasized a specific concern about hallucinations:
\begin{quote}
``\textit{This is the most important thing - if the system won't be reliable and help me with correct material, giving correct answers to my questions, why do I need it at all?}''
\end{quote}

Students described this kind of expression as harder to ignore and more difficult to reduce to a general category. While AIS responses were often experienced as composed and abstract, human interviews were experienced as more situated and consequential for how students framed the problem.

An exception was noted in Group B1, where the AIS expressed emotions such as fear, in contrast to a more authoritative human stakeholder. This suggests that while emotional expression can appear in AIS responses, students generally treated it differently from emotion grounded in lived experience.

\textbf{Cumulative sequence patterns.}
Over time, these differences accumulated into distinct patterns of use. In both sequences, students described AIS as an ``anchor'', but they used this term differently.

AIS-interview-first groups (A1–A4) generally treated AIS as an initial reference point that shaped their early understanding of the stakeholder domain. In these cases, AIS often provided a broad starting point that students later revised, narrowed, or corrected through human interviews.

Human-interview-first groups (B1–B4), by contrast, approached AIS after an initial grounding in lived experience. In this sequence, AIS was more often used as a secondary anchor, supporting validation, comparison, or expansion of insights that had already been established. Students described using AIS to revisit themes, check whether concerns recurred, or explore variations on stakeholder-derived issues.

Across both conditions, students described the AIS interview as a reference point, but its reported function varied with formal interview order. When the AIS interview preceded the human-stakeholder interview, it often served as a provisional frame that was later reconsidered. When the AIS interview followed the human-stakeholder interview, it more often supported comparison, validation, or extension of concerns already raised by the stakeholder. Because all teams constructed AIS before assignment, this contrast concerns interview order, not initial exposure to AIS.
%Across both sequences, AIS remained a reference point, but its role depended on when it entered the process. When encountered first, it tended to frame the problem space provisionally. When encountered after human interviews, it more often functioned as a tool for comparison and extension of grounded insights.

\section{Discussion}
This study examined how students used AIS alongside required human-stakeholder engagement, focusing on patterns associated with the order of the two formal interviews. All teams had constructed and evaluated an AIS before sequence assignment. In the AIS-interview-first condition, students often described the AIS interview as providing a provisional frame that was later reconsidered through the human-stakeholder interview. In the human-interview-first condition, students more often described the AIS interview as supporting comparison or validation of stakeholder-derived concerns.
These findings concern the complete sequencing conditions as implemented, including shared pre-assignment AIS preparation and discretionary opportunities for adaptation. They do not isolate a causal effect of interview order.
%This study examined how AI-Simulated Stakeholders (AIS) are used alongside human stakeholders in early-stage design work, focusing on how the order of engagement relates to students’ interpretation and use of stakeholder input. Across groups, AIS consistently served as a reference point in the design process, but its role depended on when it was introduced. When encountered first, AIS often provided an initial frame that students later revised through human interviews. When encountered after human engagement, AIS was more often used to revisit, validate, or extend insights grounded in lived experience.

%These findings suggest that AIS does not function as a neutral addition to participatory design processes. Instead, its position within the sequence of engagement is associated with different ways of interpreting stakeholder input and carrying it forward into design decisions.

\subsection{Reframing AI in Participatory Design: From Replacement to Augmentation}

Our findings speak directly to Delgado et al.'s identification of ``algorithmic proxies'' as an emerging trend in participatory AI design~\cite{delgado2023participatory}. Prior work has raised concerns that such proxies may produce ``stale modeling'' and function as a technology of de-politicization when they substitute for stakeholder participation. In contrast, our study examines AIS in settings where they are used alongside, rather than in place of, authentic stakeholder engagement.

Across groups, students used AIS alongside human-stakeholder interviews because both were required by the course. Every team had constructed, evaluated and peer-reviewed their AIS before the assignment. In the AIS-interview-first condition, the AIS interview often served as a provisional frame. In the human-interview-first condition, it more often served as a source of comparison or validation. This comparison concerns the order of the formal interviews and any discretionary adaptation between them. 
%Across groups, AIS was not used as a replacement for stakeholders, but as a reference point whose role depended on its position within the design process. When introduced early, AIS often provided a broad initial frame that students later revised through human interviews. When introduced after stakeholder engagement, it was more often used to revisit or extend insights grounded in lived experience. These patterns suggest that the effects of AIS are not determined solely by its representational quality, but also by how it is positioned relative to other sources of input over time.

At the same time, our findings are consistent with broader concerns about AI mediation. Students frequently described AIS responses as smooth, generalized, and solution-oriented, in contrast to the uncertainty, contradiction, and emotional texture of human interviews. Similar observations have been reported in prior work. Kapania et al.~\cite{kapania2025simulacrum} describe LLM outputs as lacking ``palpability'' and contextual depth, while Agnew et al.~\cite{agnew2024illusion} characterize such systems as producing an ``illusion of artificial inclusion.'' Qian et al.~\cite{qian2025second} refer to this phenomenon as ``fabricated reality,'' where representations are technically functional but lack grounding in lived experience. In our study, this appeared in the form of responses that were coherent and usable, but often required additional interpretation or validation before they could be integrated into design decisions.

These differences were also reflected in how students treated information from each source. AIS responses were often treated as provisional and subject to confirmation, while human interviews were more readily carried forward into design decisions. Rather than indicating a fixed hierarchy between sources, these patterns suggest that different types of input were treated as serving different roles within the design process. This complementarity has also been observed in collaborative ideation, where LLM-generated ideas expanded human-generated ideas and introduced additional technical and usage perspectives~\cite{shaer2024brainwriting}. However, while such expansion can be valuable when exploring a solution space, broadening the range of plausible stakeholder concerns does not necessarily improve designers' understanding of which concerns matter in practice.
Platform variation offers an alternative explanation for the Teacher System (A2) and Medical Training (A4) exceptions: groups that used non-default LLMs or configurations may have produced AIS of different specificity or coherence, independent of sequence. We did not systematically log which LLM(s) each group used for their final AIS, so this explanation cannot be ruled out and is a priority for future replication with standardized platforms.

Our findings also point to broader implications for participatory design practice. In our study, students worked with relatively accessible stakeholders, including peers and volunteers. In professional contexts, however, constraints of time, access, and resources may shape how designers choose between authentic engagement and AI-mediated representations. This raises a potential concern: AIS may be used more frequently for stakeholders who are harder to reach, including marginalized or underrepresented groups. Prior work has warned that such substitutions risk reproducing epistemic injustice by limiting whose perspectives are directly represented in design processes~\cite{fricker2007epistemic,moser2012revisiting,guan2024leveraging}.

While our study does not examine such deployments directly, the patterns we observed suggest the importance of considering how AIS is used in relation to stakeholder access. If AIS is used primarily where authentic engagement is difficult, it may inadvertently shift influence away from those stakeholders. Future work should examine how AIS is deployed in practice and how its use can be structured to support, rather than displace, engagement with underrepresented voices.

\subsection{How Students Described AIS and Empathic Engagement}

Traditional persona research positions empathy as a central mechanism in design, developed through engaging with stakeholder perspectives and synthesizing user data~\cite{nielsen2019persona,grudin2006personas}. Our findings point to a tension in how AIS contributes to this process. Students who interviewed AIS first often described using the AIS to test and reshape questions and to organize potential pain points before meeting human stakeholders. At the same time, they described AIS interactions as ``not grounded in reality'' and lacking the ``messy truths and emotion'' that characterized human interviews. This resonates with Kaate et al.'s observation of a ``fourth wall'' effect in chat-based personas, where users experience the interaction as a representation rather than as a person~\cite{kaate2025fourth}.

One possible explanation for this tension relates to how AIS was constructed and used. In traditional persona creation, designers build empathy through the process of synthesizing user data, actively constructing and refining mental models of stakeholders~\cite{grudin2006personas,nielsen2019persona}. In our study, students generated AIS through prompting, engaging primarily with the outputs rather than the full process of construction. This shift may help explain why AIS supported structured exploration of stakeholder issues while being experienced as less connected to lived experience. We do not claim that AIS reduces empathy, but that it appears to support a different mode of engagement, one that emphasizes coverage and organization over immersion in specific contexts.

The selection dynamics described by \c{S}eng\"{u}n et al.~\cite{csengun2024there} provide a useful lens for interpreting how students worked with multiple sources. While their work focuses on how designers choose among personas, students in our study made similar judgments across sources: which pain points to prioritize, which perspectives to trust, and which insights to carry forward. 
%The patterns we observed 
Table~\ref{tbl:designpoints} shows that a larger proportion of prominent human-stakeholder-derived pain points had a researcher-identified correspondence in the final designs than did prominent AIS-derived pain points. This difference concerns the frequency of coded correspondence. It does not establish the relative decision weight assigned to each source, because sources produced different numbers of concerns and the count treats an interface adjustment and a conceptual redesign as one unit each.
%suggest that insights from human stakeholders were more frequently incorporated into final designs, indicating that different sources were not only interpreted differently but also assigned different weight in decision making.

Students’ accounts also highlight differences in how emotional content was encountered and used. Human interviews were described as involving contradiction, hesitation, and strong emotional expression, which students had to interpret and integrate. AIS responses, even when they included affective language, were more often described as composed and generalized. Prior work has similarly noted that AI-generated personas may lack the experiential grounding needed to convey context and emotional nuance~\cite{matthews2012designers,kaate2025fourth}. In our study, these differences were reflected in how students described the two sources rather than in direct measures of empathic engagement. The distinction between perceived quality and human origin has also emerged in other domains. In fanfiction communities, readers and writers expressed a strong preference for human-generated content, valuing its emotional authenticity despite uncertainty about their ability to distinguish human- from AI-generated writing~\cite{alfassi2026fanfiction}. This suggests that the value attributed to a contribution may depend not only on its content but also on the human experience from which it originates.

These patterns connect to broader concerns in human–AI interaction. Qian et al.~\cite{qian2025second} describe how AI systems that ``curate human experiences'' may improve efficiency while altering users’ relationship to those experiences. In a similar way, AIS in our study appeared to support structured preparation and exploration, while human interviews introduced complexity, contradiction, and emotional weight. Rather than treating AIS as a substitute for stakeholder perspectives, these findings suggest that it may be more productive to consider how different forms of representation contribute differently to design reasoning.

Finally, some students described the contrast between AIS and human stakeholders as clarifying rather than diminishing the role of empathy. The relative simplicity and coherence of AIS responses made the complexity of human accounts more salient. %In this sense, AIS did not replace engagement with stakeholders, but in some cases functioned as a foil that highlighted the value of perspectives grounded in lived experience. --> i think this last paragraph is redundant 

\subsection{Cognitive and Pedagogical Implications of Sequential Engagement}

Students’ reflections and interview practices revealed differences in the kinds of effort and interpretation required when engaging with AIS versus human stakeholders. In particular, students described AIS interactions as requiring additional effort to extract usable insights, noting they ``had to constantly narrow and feed the AI specifics to get anything precise.'' Prior work has similarly observed that AI-generated personas may produce verbose or overly polished responses that require users to filter and interpret output carefully~\cite{kaate2025fourth}. In students’ accounts, this process involved narrowing questions, filtering AIS output, and  exploring potential issues before the human interview, but was also experienced as less connected to the situated and emotional aspects of stakeholder experience. The perceived effort of working with AI may itself shape whether designers choose to engage with it. In a study of AI use in design education, students sought AI assistance when developing interview protocols but largely refrained from using it when examining how stakeholder interviews had reshaped their understanding of the system~\cite{dines2026explain}. Together, these observations suggest that the perceived value of AI assistance depends on the kind of understanding that designers seek to develop or communicate. Importantly, additional cognitive effort in AI-supported design activities need not be detrimental to learning. Recent research on triadic collaboration among students, AI, and instructors found that instructor involvement increased germane cognitive load without a significant increase in extraneous cognitive load or team workload~\cite{triadic2026collaboration}. This raises the question of how design education might distinguish effort devoted to understanding and critically evaluating stakeholder perspectives from effort spent managing AI interactions.

Trust also differed in students’ accounts of the two sources. Several groups described not trusting AI responses ``until they're corroborated by a real interview'', while human interviews were described as immediately credible. These differences were reflected in how insights were carried forward into design decisions. Pooled across the eight groups, a larger proportion of prominent human-stakeholder-derived pain points had identified final-design correspondences than did prominent AIS-derived pain points (39/43, 90.7\%, versus 22/32, 68.8\%). At the team level, the human-derived rate was higher in four teams, equal in three, and lower in one team (A3). Because 12 prominent pain points were raised by both sources and therefore appear in both source totals, these source-specific descriptive rates overlap. They should not be interpreted as independent comparisons or as evidence of source superiority, decision weight, or causal influence. Rather than indicating a fixed hierarchy, these patterns suggest that students treated different sources as serving different roles, with human input more often grounding final decisions. This is consistent with prior work showing that limitations in AI-generated personas can affect perceived credibility and require additional validation~\cite{kaate2025you}, and with the emphasis in participatory design on engagement with authentic stakeholder voices~\cite{schuler1993participatory}.

These findings also have implications for design education. Beyond teaching students how to create and interview stakeholders, there is a need to make explicit how decisions are made about which voices are prioritized in the design process. \c{S}eng\"{u}n et al.~\cite{csengun2024there} show that persona selection is influenced by factors such as identification and perceived relevance, which can lead designers to privilege familiar perspectives. In our setting, similar dynamics appeared in how students chose which pain points to carry forward into their designs. Instructional approaches could therefore make these choices more explicit, for example by requiring students to document selection rationales, justify which perspectives are included or excluded, and reflect on how different sources influenced their final designs. Such practices align with design justice principles~\cite{costanzachock2020design} by foregrounding the question of whose experiences are treated as design-relevant.

Finally, the patterns observed here suggest that sequencing may play a role in how students develop strategies for working with AI-generated input. Students who encountered AIS early often described learning how to manage its outputs, treating them as provisional reference points while awaiting human validation. This resonates with Anthis et al.'s characterization of LLM simulations as ``silicon samples'' that can support early exploration but require critical evaluation~\cite{anthis2025llm}. Our findings extend this perspective by showing how such strategies emerge through practice when students engage with both AIS and human stakeholders over time. 

\subsection{Methodological Contributions and Theoretical Implications}

Methodologically, this study contributes by examining AIS within an ongoing design process rather than through isolated or one-time evaluations. Prior work has often compared AI- and human-generated personas in more bounded settings~\cite{goel2023preparing,smrke2025exploring}, whereas our naturalistic approach makes it possible to examine how AIS is taken up over time alongside authentic stakeholder engagement. In particular, the study highlights sequencing as an important dimension of hybrid human--AI design workflows. By following how students engaged with AIS and human stakeholders across a semester-long project, we were able to trace how different sources of input were interpreted, revised, and incorporated into design decisions.

Theoretically, our findings suggest that AIS is not well described only as an ``algorithmic proxy''~\cite{delgado2023participatory}. In our study, AIS more often functioned as a reference point within the design process: sometimes as an initial frame, and sometimes as a secondary source used for validation, comparison, or extension. This shifts the emphasis from whether AIS stands in for a stakeholder to how its position within the implemented workflow is associated with students’ interpretation of other stakeholder input over time.

This perspective also speaks to broader concerns in participatory design and design justice. Costanza-Chock argues that tools designed around efficiency can reproduce extractive dynamics when they displace meaningful engagement with affected communities~\cite{costanzachock2020design}. Our findings do not show such displacement directly, since students in this study engaged with both AIS and human stakeholders. However, they do suggest that AIS and human input were treated differently: AIS responses were more often experienced as coherent and generalized, while human interviews were treated as more grounded and more consequential for final design decisions. These patterns underscore the importance of examining not only whether AI-generated stakeholder representations are used, but how they are positioned relative to authentic voices in the design process.

\subsection{Design Implications for Conversational AIS}

Our findings suggest several directions for the design of conversational AIS as tools for participatory design. The recommendations below vary in evidentiary grounding: some follow directly from patterns we observed (e.g., students' need to validate AIS before trusting it), while others are speculative directions motivated by, but not directly tested in, our data (e.g., productive-friction features).

\textit{Clarifying function and boundaries.} Prior work on algorithmic proxies has raised epistemological concerns about conflating synthetic outputs with lived experience~\cite{kaate2025fourth,kapania2025simulacrum}. In our study, students often treated AIS responses as provisional and sought confirmation through human interviews. This suggests a need for AIS to make their representational limits more visible. Rather than presenting as authentic stakeholder voices, conversational AIS may benefit from explicitly communicating their role as exploratory or provisional representations. This could take the form of system-level transparency features such as: (a) uncertainty markers (e.g., hedging language, confidence indicators, or statements such as ``this is a generalization, not based on lived experience''); (b) scope disclaimers (e.g., notes that the AIS reflects patterns from available data rather than a specific individual or community); and (c) indicators when responses rely on common or stereotypical patterns. Such features may help designers interpret AIS outputs as starting points that require validation through stakeholder engagement.

\textit{Designing for productive friction.} Unlike static personas, which require designers to engage in synthesis and interpretation~\cite{grudin2006personas,marsden2016stereotypes}, conversational AIS can present coherent and readily usable narratives. In our study, this sometimes reduced the need to work through ambiguity. One possible design direction is to introduce forms of ``productive friction,'' such as hesitation, partial answers, or conflicting perspectives. These features would not aim to simulate authenticity, but to make visible the limits of the representation and encourage further questioning. This aligns with Anthis et al.'s argument that simulations are most useful when they surface possibilities rather than provide definitive answers~\cite{anthis2025llm}.

\textit{Adaptive support across use.} The differences between AIS-interview-first and human-interview-first sequences suggest that AIS may play different roles over time. Early interactions could support exploration, for example by helping designers generate questions or map a broad set of potential concerns. Later interactions could foreground limitations by surfacing uncertainty, gaps, or inconsistencies. Explicitly prompting designers to consider what may be missing from AIS responses could further encourage critical engagement, drawing on evidence that prompting supports the recognition of missing information~\cite{benshoshan2026absence}. Such a shift may help direct designers back toward authentic stakeholder engagement rather than allowing AIS to stand in for it. This perspective resonates with Kaate et al.'s observation~\cite{kaate2025fourth} that different persona formats involve different trade-offs in use.

\textit{Integration into participatory workflows.} Finally, AIS design may benefit from extending beyond one-on-one interaction to support integration with broader participatory processes. For example, systems could support comparison between AIS responses and stakeholder data, or help teams track where synthetic and human inputs converge or diverge. Such features could position AIS more clearly as part of a larger design workflow, supporting augmentation rather than substitution.

\textit{Summary.} Taken together, these implications highlight a core design challenge: conversational AIS need not aim to replicate the authenticity of lived voices, but rather to make their synthetic character visible and usable. Designing AIS that balances transparency, friction, and evolving support may help them complement, rather than compete with, stakeholder participation.

\subsection{Implications for Education}

Building on Goel et al.’s call to integrate AI persona creation into curricula to prepare future UX designers~\cite{goel2023preparing}, we consider what our findings suggest for design education.

Students in our study encountered both the affordances and limitations of AI Simulated Stakeholders (AIS), navigating the tension between polished but generalized synthetic responses and the more complex, sometimes contradictory accounts provided by human stakeholders. These experiences point to several pedagogical opportunities.

First, AIS can serve as low-a stakes practice environments. Interviewing an AIS allowed students to rehearse questioning techniques, refine interview protocols, and reduce the anxiety associated with initial stakeholder engagement. In this sense, AIS functioned less as stand-ins for authentic stakeholders and more as tools for developing confidence and procedural fluency.

Second, the contrast between AIS responses and human accounts was repeatedly described by students as informative. The generality and solution-oriented nature of AIS responses made the specificity and emotional expression of human interviews more visible. Learning activities that intentionally juxtapose synthetic and authentic perspectives may therefore help students attend more closely to nuance, context, and variation in stakeholder accounts.

Third, our findings highlight the importance of integrating structured reflection into the use of AIS. Without explicit prompts to question the limits of AIS representations, some students reported a tendency to treat responses as sufficiently informative or to rely on their convenience. Educational approaches could therefore include activities that require students to reflect on issues such as bias, representational limits, and the distinction between augmentation and substitution.

Taken together, these observations suggest that AIS can function as provisional reference points in the learning process, shaping how students prepare for and reflect on engagement with human stakeholders. Integrating AIS into design curricula may support the development of interviewing skills, analytic practices, and critical reflection, while reinforcing the role of authentic stakeholder voices in participatory design.

\subsection{Limitations and Future Directions}
\label{sec:limit}

This study was designed as an ecologically grounded investigation of how students incorporated AI Simulated Stakeholders (AIS) into authentic project work. Its ecological validity is important because actual uses of generative AI involve variation in project domains, available systems, prompting strategies, and opportunities for revision. Allowing teams to make these choices provided insight into how AIS may be appropriated in practice. 

At the same time, this naturalistic design places limitations on the interpretation of the findings.
First, the study involved 21 novice designers, 19 undergraduate and two graduate students, in a single academic course. The findings are therefore most directly applicable to similar educational settings, and their transferability to professional or other educational contexts requires further examination.  While all students completed the same preparatory activities, we did not administer a pre-course measure of their prior experience with design methods, generative AI, personas, or stakeholder interviews. We therefore cannot determine how differences in prior experience may have influenced their construction, evaluation, or use of AIS.
Second, the study followed eight teams during one course project. It does not establish whether the practices students developed persisted over time or transferred to later design work. The number of participants and teams was also appropriate for examining qualitative patterns but not for making statistical population-level claims.
Third, sequence was intertwined with opportunities for AIS refinement. Human-interview-first teams could incorporate insights from human stakeholders when refining their AIS. This reflects how sequencing may operate in practice and is therefore substantively relevant.  However, it also means that the comparison captures both interview order and the information available during AIS development. Similarly, the common AIS-evaluation framework described in Section 4.1 provided teams with a shared vocabulary but did not ensure equivalent AIS quality, prompting practices, participant experience, or prior expertise. The observed differences should therefore be interpreted as patterns associated with the complete sequencing conditions rather than effects of interview order alone. All teams had constructed, evaluated, and peer-reviewed AIS before sequence assignment. The study therefore does not compare initial exposure to AIS with introduction after human engagement. It compares the order of the two formal interviews after shared AIS preparation. That shared preparation may already have influenced how teams understood their stakeholders in both conditions.
Fourth, teams adapted AIS to their projects rather than following a fully standardized implementation. Most used ChatGPT, generally through its free web interface, while some used or tested other large language models. Prompt content, background information, interaction histories, and revision practices also varied. This variation strengthens the study’s account of how students appropriate AIS under realistic conditions, but it limits our ability to attribute particular findings to a specific model or interaction feature. In several teams, the AIS and human interviewees occupied related but not identical roles or experience profiles. Observed source differences may therefore partly reflect differences in stakeholder role, specificity, or perspective rather than source type alone. In addition, prominence was inferred from repetition or expressed intensity. Source-specific communication styles, such as repetitive AIS output or emotionally emphatic human accounts, may therefore have affected which concerns were classified as prominent.
Fifth, teams selected different project domains and stakeholder groups (Table 1). This diversity allowed the study to examine AIS across varied design situations, but differences in emotional sensitivity, stakeholder expertise, technical complexity, and stakeholder access may also have shaped teams’ experiences. These domain characteristics cannot be separated fully from the sequencing conditions in the present design.
Sixth, the qualitative analysis has limitations. Coding was conducted with knowledge of the completed project materials, creating some possibility of hindsight bias. Nevertheless, the analysis identified multiple pain points that did not subsequently appear in teams’ reports or designs, suggesting that the coding was not reconstructed solely from final outcomes. The post-course evidence was also selective. Only three of the 21 students participated approximately four months after the course, and all three represented AIS-interview-first teams. Their accounts may be affected by selective participation and delayed recall and cannot support cross-condition conclusions. Findings based on the broader course-embedded corpus are therefore distinguished from observations supported only by the post-course discussion or individual interview.
Finally, the analysis focused on how students described their cognitive work and relational considerations when engaging with AIS and human stakeholders. It did not independently assess downstream design quality, implementation success, or stakeholder satisfaction. The findings therefore illuminate students’ processes and interpretations but do not establish the comparative effectiveness of the resulting designs.

These limitations point to opportunities for complementary research. Longitudinal and multi-site studies could examine whether AIS-related practices persist beyond a single project and extend to professional or community-based settings. More controlled studies could separately examine interview order, AIS refinement, prior experience, model choice, and project domain. Such studies would complement the ecological perspective developed here. 
Future evaluations should also examine design quality and stakeholder experience directly, particularly in work involving underrepresented or difficult-to-reach communities, where substitution or misrepresentation may have especially serious consequences.
\section{Conclusions}

This paper examined how students combined AIS and human-stakeholder input when both forms of engagement were required. All teams constructed and evaluated AIS before assignment and then conducted the two formal interviews in different orders. Students in the AIS-interview-first condition often described the AIS interview as a provisional frame that was later reconsidered through human engagement. Students in the human-interview-first condition more often described the AIS interview as supporting comparison or validation. These are exploratory patterns associated with the complete conditions as implemented, not isolated effects of interview order.

By situating AIS within related approaches such as algorithmic proxies, personas, and LLM-based simulations, this paper contributes both conceptual clarification and empirical observations. The findings suggest that AIS can support preparation and procedural skill development, while also tending to omit the hesitation, contradiction, and affect that characterized human stakeholder accounts. These differences help explain how students positioned AIS and human inputs within their design process.

The study does not establish whether designers would retain human participation if it were optional, particularly under constraints of time, resources, or stakeholder access. It shows instead how students combined AIS and human stakeholder input when both were required. Within that setting, AIS supported activities such as rehearsal, provisional framing, comparison, and validation, while human interviews provided situated and emotionally textured accounts. These findings motivate treating AIS as provisional complements within educational workflows. Whether AIS supplements or displaces human participation in unconstrained practice remains an open empirical question.

\section*{Ethical approval}
This research was conducted in compliance with ethical guidelines and received approval from the Institutional Review Board (IRB) of the University of Haifa (approval 450/25) 
\section*{Disclosure statement}
The authors declare no conflicts of interest.
\section*{Funding}
The authors received no financial support for the research, authorship, or publication of this article.
\section*{Data availability statement}
Deidentified coding materials and representative quotations are provided in the appendices. 
\section*{Author Contributions}
OM: Conceptualization, design, data acquisition, data analysis, paper writing, paper review; OA: data acquisition, data analysis, paper writing, paper review; ND: Data analysis; OS: Conceptualization, paper  writing, paper review. 
\bibliographystyle{ACM-Reference-Format}
\bibliography{cites}
\clearpage
\appendix

\section{The code book for one of the groups}
Table~\ref{tab:caretech-codebook} presents the complete CareTech codebook used as the detailed example of pain-point identification, prominence classification, and design correspondence.
\label{sec:apdxA}
{\footnotesize
\setlength{\tabcolsep}{3pt}
\begin{longtable}{@{}
  P{1.2cm}   % Pain Point
  P{1.3cm}   % Prominence
  P{3.4cm}   % Description
  P{3.4cm}   % Representative Quote
  P{1.4cm}   % Reason (Prominence)
  P{1cm}          % Influence Level
  P{1.4cm}   % Design Component
@{}}
 
\caption{CareTech Codebook --- Pain Point Identification, Prominence Classification, and Design Tracing}
\label{tab:caretech-codebook}\\
 
\toprule
\textbf{Pain Point} &
%\textbf{Source} &
\textbf{Prominence} &
\textbf{Description} &
\textbf{Representative Quote} &
\textbf{Reason (Prominence)} &
\textbf{Corr. category} &
\textbf{Design Component} \\
\midrule
\endfirsthead
 
\multicolumn{7}{c}{\textit{(continued from previous page)}}\\[2pt]
\toprule
\textbf{Pain Point} &
%\textbf{Source} &
\textbf{Prominence} &
\textbf{Description} &
\textbf{Representative Quote} &
\textbf{Reason (Prominence)} &
\textbf{Corr. Category} &
\textbf{Design Component} \\
\midrule
\endhead
 
\midrule
\multicolumn{7}{r}{\textit{(continued on next page)}}\\
\endfoot
 
\bottomrule
\endlastfoot
 
% --- human stakeholder rows ---
\multicolumn{7}{l}{\textit{\textbf{Source: human stakeholder}}}\\
\midrule
 
Poor information system & Prominent &
that The organization's non-profit status means computer systems and software are not advanced. &
``One of the serious problems in the organization's operations is its poor computer system \ldots{} there are much better off-the-shelf programs than what we have. But we're a nonprofit, so we don't have a lot of money.'' &
Frustration; Repeats &
2 &
--- \\[4pt]
 
Toll of summarizing conversations & Prominent &
Summarizing conversations demands significant effort and time, and affects the volunteer's ability to reset for the next call. &
``Because a conversation ends, and then it has to be concluded \ldots{} you have to take a break \ldots{} You need to get out of the conversation. You need to maybe reset, maybe relax.'' &
Frustration; Repeats &
4 &
Automatic call summarization \\[4pt]
 
Payment for training  & Mentioned &
Volunteers pay out-of-pocket for the training course. &
``It cost money. I mean, I paid money to attend the course \ldots{} I paid 500 [currency] to attend the course.'' &
Mentioned casually &
1 &
--- \\[4pt]
 
Training can be insufficient  & Mentioned &
Some volunteers finish training and are still not ready to handle calls independently. &
``There are volunteers who struggle alone even after they have supposedly already been trained.'' &
Mentioned casually &
1 &
--- \\[4pt]
 
Feedback  & Prominent &
Volunteers place high importance on receiving regular, specific feedback about their calls. &
``Even if they tell [you], listen, you didn't do that well here. It also gives a good feeling. You are being treated. Because you can't just volunteer in space.'' &
Repeats; Social aspect &
4 &
Volunteer Feedback System \\[4pt]
 
Sudden call disconnections & Prominent &
Calls drop without warning and it is not possible to call the caller back. &
``\ldots{}calls sometimes drop without warning\ldots'' &
Frustration; Repeats; Social aspect; Technical failure &
4 &
Enhanced disconnection handling \\[4pt]
 
Anonymity and privacy & Prominent &
Preserving anonymity allows callers to speak openly without fear of exposure. &
``Anonymity, it must be preserved, and it is sacred. Because otherwise people won't call, people trust it, that it's anonymous.'' &
Ethics; Social aspect; Regulatory/Legal &
4 &
Temporary anonymous profile \\[4pt]
 
Long calls  & Mentioned &
Some calls are longer than usual and require more emotional resources from volunteers. &
``And sometimes you need to talk more, based on interest, based on the feel of the applicant.'' &
Tolerated; Seldom &
1 &
--- \\[4pt]
 
Disrespectful discourse  & Prominent &
Some callers address volunteers in a disrespectful manner. &
``If someone speaks a language I'm not willing to hear, I say, I'm not willing to hear it, I hang up the call.'' &
Repeats; Social aspect; Emotional aspect &
1 &
--- \\[4pt]
 
Emergency protocol  & Prominent &
Emergency protocol requires phone escalation to management and may be too slow to address crises. &
``I'll call whoever's in charge of me. The branch manager \ldots{} Yes, and they have some way \ldots{} Call the police.'' &
Ethics; Standard protocol; Regulatory/Legal &
4 &
Emergency response protocol \\[4pt]
 
Emotional challenge  & Mentioned &
Some calls have difficult content requiring processing, though support is available within the organization. &
``There are conversations like this that \ldots{} I think about them, I remember them, but I also have someone to talk to about them (within the organization).'' &
Bearable; Already resolved; Seldom &
1 &
--- \\[4pt]
 
\midrule
\multicolumn{7}{l}{\textit{\textbf{Source: AI Simulated Stakeholder (AIS)}}}\\
\midrule
 
Emotional challenge & Prominent &
Calls with difficult content require containment and self-protection; certain periods increase mental load. &
``There are moments when it is very difficult to hear and cope with the pain of others, especially when you feel that you yourself are going through a difficult time.'' &
Emotional aspect; Cognitive load &
1 &
--- \\[4pt]
 
Sharing among peers & Mentioned &
Informal peer guidance by experienced volunteers meaningfully supports newcomers, complementing formal training. &
``I think this informal guidance is just as important as formal training because it gives new volunteers a sense of security and direct support from peers.'' &
Mentioned casually &
1 &
--- \\[4pt]
 
Technical multitasking during conversations  & Mentioned &
Managing conversation content alongside technical tasks creates pressure and reduces focus on callers. &
``All of these tools are supposed to reduce stress and allow me to be more focused on responding to the caller, and not just on the technical process.'' &
Mentioned casually &
1 &
--- \\[4pt]
 
Workload volume &
 Prominent &
High call volumes or long calls create workload pressure that risks reducing service availability. &
``When there are many calls at the same time, we \ldots{} try to ensure that everyone gets the time they need, but also that conversations do not extend in a way that prevents volunteers from serving all callers.'' &
Repeats; Emotional aspect; Cognitive load &
2 &
--- \\
 
\end{longtable}
}
 
\vspace{6pt}
{\footnotesize\textit{Note.} ``Corr. Category'' = Correspondence/attribution category (scale 1--4). Category 1 = no identified design correspondence; 2 = contextual reference; 3 = researcher-inferred feature correspondence; 4 = student-reported foundational influence.
Source ``AIS'' = AI Simulated Stakeholder.
Entries marked ``---'' had no design component specified.}

\section{Case Study: Elderly Transportation Technology Group (B1)}
\label{ref:apdx}

%\subsection{Case Study: Elderly Transportation Technology Group (B1)}
To illustrate our analytic process and the kinds of patterns it revealed, we present one project in detail: \textit{Elderly Transportation Technology}. Group B1 focused on redesigning a public transportation application for elderly users, drawing on both a human stakeholder interview and an AIS interview. The group’s final design included six components, five of which had researcher-identified correspondences with concerns raised by both sources: voice commands, configurable text and buttons, augmented reality navigation assistance, step-by-step navigation, and improved update rate and validation (Table~\ref{tbl:B1}). One component - audio alerts for route changes - had an identified correspondence only with a concern raised in the AIS interview. This source correspondence does not establish that the AIS statement caused the component.

\begin{table}[h]
\centering
\caption{Detailed example of design components and corresponding interview sources for Group B1, Elderly Transportation Technology}
\begin{tabular}{p{.5cm} p{5cm} p{6cm} p{2cm}}
%\hline
\textbf{\#}& \textbf{Design Idea} & \textbf{Design Component} & \textbf{Corresponding Interview Source(s)} \\ \hline
1 &A clear microphone button allowing hands-free operation 
& Voice commands (microphone button) 
&  Both \\ \hline
2 &Clear text and buttons designed for users with visual impairments 
& Size configurable buttons and  text 
& Both \\ \hline
3 &Help identify the station and direction 
& AR-enabled smart navigation 
& Both \\ \hline
4 &Reduce confusion and provide real-time updates 
& Audio alerts 
& AIS \\ \hline
5 &Reduced visual load 
& Step-by-step navigation, each step on a separate screen
& Both\\ \hline
6 &Improved information reliability 
& Improve update rate and validation, especially in cases of delays and cancellations 
& Both \\ %\hline
\end{tabular}
\label{tbl:B1}
\end{table}

Table~\ref{tbl:ElderPP} illustrates how pain points were identified and traced into the final design. For example, the AIS described anxiety during route changes, and this concern was reflected in the addition of audio alerts. Because this concern recurred across the AIS interview and was central to multiple responses, it was coded as \textit{prominent}. In the final design, a corresponding feature-level component was identified in the final design. Because the team did not explicitly state that this AIS concern prompted the component, the case was coded as a researcher-inferred feature correspondence rather than as student-attributed influence.

\begin{table}[htbp]
\scriptsize
\setlength{\tabcolsep}{2pt}
\renewcommand{\arraystretch}{1.15}
\centering
\caption{B1 Elderly Transportation Technology Group: Coding of how insights from interviews were reflected in the design, including source, rationale, and degree of incorporation into the final design.}
\label{tbl:ElderPP}

\begin{tabular}{@{}%
p{0.2\linewidth}%
>{\centering\arraybackslash}p{0.08\linewidth}%
p{0.56\linewidth}%
>{\centering\arraybackslash}p{0.14\linewidth}@{}}
\toprule
\textbf{Prominent Pain Point}
& \textbf{Source}
& \textbf{Relevant Quote}
& \makecell{\textbf{Design correspondence}\\
            \textbf{and intervention level}\\
            \textbf{(see Table~\ref{tbl:pp-criteria})}} \\
\midrule

\multirow{2}{=}{Sense of independence and security}
& Human
& \textit{At first I needed help from my grandchildren to understand
how to plan a route \ldots{} I already know the regular stations, and
this gives me confidence.}
& \multirow{2}{=}{\textit{Contextual reference}; no intervention level assigned} \\

& AIS
& \textit{I love that it gives me independence. I can leave whenever I
want and do not need to ask my children to drive me. It is a good
feeling to know that I can manage on my own.}
& \\
\midrule

\multirow{2}{=}{Dense text and tiny buttons hinder readability and accuracy}
& Human
& \textit{The hardest part was reading the small text and figuring out
which route was most suitable.}
& \multirow{2}{=}{\textit{Student-reported foundational influence};
intervention level 1} \\

& AIS
& \textit{The text is small, and I have difficulty reading it, even
with glasses \ldots{} If there were an option to see fewer details and
enlarge the buttons, it would be much more convenient.}
& \\
\midrule

\multirow{2}{=}{Visual clutter and improperly presented information create confusion and uncertainty}
& Human
& \textit{Sometimes there are too many options, and it is confusing
\ldots{} There is a lot of information all at once, and it makes it
hard for me to understand what to do.}
& \multirow{2}{=}{\textit{Researcher-inferred feature correspondence};
intervention level 2\textsuperscript{*}} \\

& AIS
& \textit{There are too many details on the screen at once, and it
confuses me. I do not always know what is really important.}
& \\
\midrule

\multirow{2}{=}{Difficulties navigating with the application}
& Human
& \textit{It is not very clear to me. The routes are displayed in a
complicated way, and sometimes I am not sure which route is the best.}
& \multirow{2}{=}{\textit{Student-reported foundational influence};
intervention level 3} \\

& AIS
& \textit{I do not think it is organized enough. Sometimes it is hard
to understand where to go, and after a few minutes of use I no longer
remember the route. It stresses me out.}
& \\
\midrule

Stress and anxiety due to route changes
& AIS
& \textit{The information is usually reliable, but sometimes there are
delays and changes that the application does not update in time. I
think that if there were clear voice instructions announcing changes,
it would really help me feel more secure.}
& \textit{Researcher-inferred feature correspondence};
intervention level 2 \\
\bottomrule
\end{tabular}

\vspace{0.25em}
\parbox[t]{\linewidth}{%
\scriptsize
\textit{\textsuperscript{*}This pain point corresponded to the level-2
design components numbered 4 and 6 in Table~\ref{tbl:B1}.}
}
\end{table}

\begin{figure}[!h]
    \centering
    \includegraphics[width=.5\linewidth]{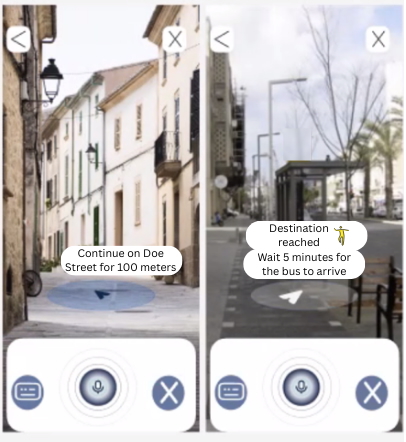}
    \caption{Group B1, Elderly Transportation Technology: Incorporating a design suggestion made by both the human stakeholder and the AIS to add an AR layer to the directions. Minor details in the picture were changed to maintain anonymity.}
    \label{fig:B1AR}
\end{figure}

Figure~\ref{fig:B1AR} shows one of the resulting design elements: an added AR layer. This feature could be traced to both the human stakeholder and the AIS, each of whom described difficulty understanding navigation and determining direction. In our coding, this design element was treated as a level 3 intervention because it altered the conceptual structure of the app rather than only adjusting an existing feature.

\section{Protocol for the post-course discussion and individual interview}
\label{sec:interviewprotocol}
% Requires: \usepackage{enumitem}
% (for \begin{enumerate}[resume] and label formatting)

\subsection*{Opening and Consent}

\noindent Hello, and thank you for coming. I am [anonymized], a faculty member in the department. In this discussion, we will talk about the teaching processes that incorporate GenAI in the systems design course.

\medskip
\noindent During this conversation, there are no right or wrong answers. Our aim is to help us refine the insights that emerged in the discussion held during the course.

\medskip
\noindent We will not discuss personal or sensitive topics. Even so, if at any point you feel uncomfortable, simply tell us---we can skip any of the questions, or end the conversation altogether.

\medskip
\noindent Everything we talk about today is confidential, and your names will remain confidential.

\medskip
\noindent May I have your permission to record this interview? The recording will be stored confidentially and used for research purposes only.

\medskip
\noindent Before we begin, do you have any questions?

\medskip
\noindent When you answer during the conversation, please state your name.

\subsection*{Part 1: Preparation and the Teaching Process}

\begin{enumerate}[leftmargin=*]
    \item Did the process you went through in class before creating the persona prepare you sufficiently?
    \item Do you have any recommendations for improving the teaching process?
\end{enumerate}

\subsection*{Part 2: Creating the Persona}

\begin{enumerate}[leftmargin=*, resume]
    \item Did the process of creating the persona contribute to your understanding of the project? In what way?
    \item As part of the creation process, you wrote several prompts before arriving at the final one. How did you decide to stop? How did you decide it was good enough?
\end{enumerate}

\subsection*{Part 3: Interviewing the Persona}

\noindent Now let us look more closely at the part in which you interviewed the persona.

\begin{enumerate}[leftmargin=*, resume]
    \item Did the interview itself contribute to you? In what way?
    \item What was the advantage of the persona, from your point of view?
    \item Was there a case in which you disagreed with the persona? What did you do?
    \item If you did disagree with the persona, did you think the problem lay in the way you had created it, or that it was simply wrong?
    \item Did you see the persona as a source of authority?
    \item Did you change your design for the system following the conversation with the persona?
    \item Did you feel that your own ability to create a persona affected how much you trusted it?
\end{enumerate}

\subsection*{Part 4: Back to the Creation Process}

\begin{enumerate}[leftmargin=*, resume]
    \item Returning to the process of creating the persona: did you use questions from the interview you had planned in order to assess whether the persona was good enough?
\end{enumerate}

\subsection*{Part 5: Closing}

\begin{enumerate}[leftmargin=*, resume]
    \item In the end, would you say this contributes to the design process? If so, how would you recommend integrating it into the design process---should the interview with the persona come before or after the interview with a human stakeholder?
\end{enumerate}

\section{Codebook for the post-course discussion, individual interview, end-of-course class discussion, and written reflections}
\label{app:qualitative-codebook}

We created a joint codebook for the two-participant post-course discussion, individual post-course interview, end-of-course class discussion, and 13 student-submitted written reflections.

\begin{table*}[t]
\caption{Codebook for the post-course discussion, individual interview, end-of-course class discussion, and written reflections, grouped into five themes. Quotations were translated from Hebrew. Within the table, [CC] denotes a contradictory case.}
\label{tab:quotation-codebook}
\scriptsize
\setlength{\tabcolsep}{3pt}
\renewcommand{\arraystretch}{1.02}
\begin{tabularx}{\textwidth}{@{}
  >{\raggedright\arraybackslash}p{0.19\textwidth}
  >{\raggedright\arraybackslash}p{0.26\textwidth}
  >{\raggedright\arraybackslash}X@{}}
\toprule
Code & Brief operational meaning & Representative quotation \\
\midrule
\multicolumn{3}{@{}l}{\textbf{Preparation and confidence}} \\
\addlinespace[1pt]
Teaching and guidance beforehand &
Whether class instruction, readings, and examples left students feeling ready to build an AIS, and what was missing. &
\emph{I like to work very systematically, so it was difficult not knowing exactly how to do it.} \\
Building it in rounds, and knowing when to stop &
Making the AIS over several attempts, and how students decided it was good enough to use. &
\emph{We had two or three interactions.} \\
Using, or dropping, the evaluation tools taught &
Whether students applied the AIS-evaluation methods from class, and whether they felt those methods worked. &
\emph{The hats we didn't really know how to use properly, so we preferred not to use them} \\
Confidence in their own prompting &
Students blaming or crediting their own prompt-writing skill for the quality of what the AIS produced. &
\emph{It is hard for me to trust the result if I do not think I made the prompt correctly.} \\
\addlinespace[2pt]
\midrule
\multicolumn{3}{@{}l}{\textbf{Where the AIS fell short}} \\
\addlinespace[1pt]
Too generic or stereotyped &
A bland, textbook version of the stakeholder, with no specifics, variation, or unusual cases. &
\emph{It was very vanilla, not entirely accurate.} \\
Focused on the wrong thing, or stayed shallow &
Latching onto an obvious, familiar issue while missing the problem the human stakeholder actually described. &
\emph{It saw the word ADHD and focused on how people with ADHD need help with concentration and maybe time management. It started talking about those things, and completely ignored the real struggles the [human] student described.} \\
Jumped to a solution that ignored what already exists &
Offering a tidy fix that showed no awareness of the tools, rules, or practices already in place in that setting. &
\emph{They already have this solution, they already have an app that gives breaks, so maybe the AI did not check deeply enough.} \\
Defended the organization instead of admitting problems &
Confident, PR-style answers that protected the company or institution rather than revealing what was wrong. &
\emph{It was very political, giving answers in line of [describing AIS 'internal perspective']: 'I don't want to damage the company's image, so I won't say anything. In fact, I'll say positive things instead.' But we were aiming to solve privacy problems in this company, and it responded: 'Such problems do not exist, and everything is fine.'} \\
Told students what they wanted to hear &
Answers that seemed aimed at pleasing the interviewer, or that simply echoed back concerns raised in the prompt. &
\emph{There also needs to be a way to make sure it does not drift too far and start trying to please you.} \\
Contradicted itself or drifted from the AIS brief &
Saying opposite things at different points, forgetting earlier answers, or inventing details never given in the brief. &
\emph{It has a really short memory. By question eight it has forgotten everything it answered before. It invents a new persona and answers as that new persona.} \\
\addlinespace[2pt]
\midrule
\multicolumn{3}{@{}l}{\textbf{Work needed to get usable answers}} \\
\addlinespace[1pt]
Answers too long to be usable &
Output so long or unfocused that students had to hunt through it for anything worth keeping. &
\emph{The interview was 23 pages of nothing.} \\
Narrowing the question to get something specific &
Adding detail or constraints to a question so the AIS would stop answering in generalities. &
\emph{had to constantly narrow and feed the AI specifics to get anything precise} \\
Fixing answers mid-conversation &
Steps students took during the interview to correct a bad or off-track answer, rather than starting over. &
\emph{I'd give the same prompt again and add a space character, and usually it would correct itself. When we saw a mismatch we'd simply instruct it: this fits less, drop that part.} \\
\addlinespace[2pt]
\midrule
\multicolumn{3}{@{}l}{\textbf{What made an account believable}} \\
\addlinespace[1pt]
Believing it only after outside confirmation &
Holding back on trusting AIS output until something else, usually a human stakeholder, confirmed it. &
\emph{I didn't really believe what ChatGPT wrote until I saw it confirmed in the real interview.} \\
Emotion and lived detail as a sign of the real thing &
Feeling, emphasis, or concrete personal experience that made a human account land differently from AIS output. &
\emph{In the real interview, you could hear the pain in her voice. ChatGPT just wrote everything in the same way.} \\
The human contradicted what the team believed &
Surprising or opposing input from a human stakeholder that unsettled what the team thought it already knew. &
\emph{The real participant said things we didn't expect and contradicted our assumptions.} \\
AIS and human pointed to the same issue &
The same problem coming up in both the AIS and the human interview, which raised students' confidence in it. &
\emph{It sharpened the problems that the human stakeholder had also mentioned.} \\
\addlinespace[2pt]
\midrule
\multicolumn{3}{@{}l}{\textbf{What the AIS was useful for, and when}} \\
\addlinespace[1pt]
Practicing the interview and testing the questions &
Using the AIS to try out the interview guide, get a feel for the interviewee, or calm nerves before meeting a human stakeholder. &
\emph{broke the fear} \\
Sharpening which problem to work on &
Building or interviewing the AIS helping the team narrow down the problem itself, separate from any solution it suggested. &
\emph{It didn't help me with the final solution I came up with, but in the end it simply focused the problem.} \\
Whether it changed the final design &
Whether AIS input actually reached the finished design, sent the team off course, or took time without adding anything. &
\emph{Unfortunately, in our case, it didn't contribute to the solution at all and actually delayed us. \newline [CC]: ``At first our system was more focused on load in terms of amount of work... Afterwards we saw that the load is more on the emotional side, so we decided the system would be one that advises them to take breaks... And that was because of the persona, not because of the person.''} \\
Which order students preferred, and why &
A stated preference for AIS-interview-first or human-interview-first, together with the reason given for it. &
\emph{It is better to try the models as a foundation and build on that according to the human, rather than take the human and then try to apply that to the model.} \\
\addlinespace[2pt]
\bottomrule
\end{tabularx}
\end{table*}
%\appendix
\FloatBarrier

\section*{Author biographies}

\noindent\textbf{Osnat Mokryn} is an Associate Professor of Information Systems at the University of Haifa, where she heads the HCI master's program. She studies how people keep their agency and judgment when working with AI, from design and decision-making to high-stakes teamwork. She also chairs the ACM IUI Steering Committee.

\medskip
\noindent\textbf{Oshrat Ayalon} is a Senior Lecturer (Assistant Professor) in the Information Systems Department at the University of Haifa. Her research lies at the intersection of human-computer interaction and information privacy and security, promoting ethical and safe technologies. She holds a Ph.D. and M.Sc. in Industrial Engineering from Tel Aviv University.   % up to 50 words

\medskip
\noindent\textbf{Noga Dines} is an M.Sc. student in the Department of Information Systems at the University of Haifa, specializing in Human Computer Interaction. She holds a B.Sc. in Information Systems and serves as a research and teaching assistant. Her research examines human AI teaming, interactional authority, and design education.

 % up to 50 words

\medskip
\noindent\textbf{Orit Shaer} is Associate Provost for Wellesley in the World, the Michael and Denise Kellen '68 Chair in the Sciences, and Professor of Computer Science at Wellesley College. A 2026–27 Radcliffe Fellow, she studies tangible and embodied interaction and human-AI creative collaboration, asking how emerging technology can support human flourishing. % up to 50 words

\end{document}